\documentclass[manuscript]{aastex61}
\usepackage{amsmath,amstext,amssymb}
\usepackage[T1]{fontenc}
\usepackage{apjfonts} 
\usepackage{color,soul}
\usepackage[figure,figure*]{hypcap}
\usepackage[toc,acronym]{glossaries} 
\usepackage{glossary-longbooktabs}
\usepackage{ulem}
\usepackage{pgf,tikz}
\usepackage{mathrsfs}
\usepackage{pstricks-add}
\usetikzlibrary{arrows}

\newacronym{AU}{AU}{Astronomical Unit [1.5e11 m]}  
\newacronym{pc}{pc}{parsec}
\newacronym{mas}{mas}{milliarcsecond}
\newacronym{nm}{nm}{Nanometer}
\newacronym{CTE}{CTE}{coefficient of thermal expansion}

\newacronym{smc}{SMC}{Small Magellanic Cloud}
\newacronym{lmc}{LMC}{Large Magellanic Cloud}
\newacronym{ism}{ISM}{interstellar medium}
\newacronym{mw}{MW}{Milky Way}
\newacronym{epseri}{$\epsilon$ Eri}{Epsilon Eridani}
\newacronym{EKB}{EKB}{Edgeworth-Kuiper Belt}

\newacronym{CFR}{CFR}{Complete Frequency Redistribution}

\newacronym{nasa}{NASA}{National Aeronautics and Space Agency}
\newacronym{esa}{ESA}{European Space Agency}
\newacronym{omi}{OMI}{\textit{Optical Mechanics Inc.}}
\newacronym{gsfc}{GSFC}{\gls{nasa} Goddard Space Flight Center}
\newacronym{stsci}{STScI}{Space Telescope Science Institute}
\newacronym{nsroc}{NSROC}{\gls{nasa} Sounding Rocket Operations Contract}
\newacronym{wff}{WFF}{\gls{nasa} Wallops Flight Facility}
\newacronym{wsmr}{WSMR}{White Sands Missile Range}

\newacronym{irac}{IRAC}{Infrared Array Camera}
\newacronym[plural=CCDs, firstplural=charge-coupled devices (CCDs)]{ccd}{CCD}{charge-coupled device}
\newacronym[plural=EMCCDs, firstplural=electron multiplying charge-coupled devices (EMCCDs)]{EMCCD}{EMCCD}{electron multiplying charge-coupled device}

\newacronym{DM}{DM}{Deformable Mirror}
\newacronym{MCP}{MCP}{ Microchannel Plate }
\newacronym{ipc}{IPC}{Image Proportional Counter}
\newacronym{cots}{COTS}{Commercial Off-The-Shelf}
\newacronym{ISR}{ISR}{Incoherent Scatter Radar }
\newacronym{atcamera}{AT}{Angle Tracker}
\newacronym{MEMS}{MEMS}{microelectromechanical systems}
\newacronym{QE}{QE}{quantum efficiency}
\newacronym{RTD}{RTD}{Resistance Temperature Detector}
\newacronym{PID}{PID}{Proportional-Integral-Derivative}
\newacronym{PRNU}{PRNU}{photo response non-uniformity}
\newacronym{DSNU}{PRNU}{dark signal non-uniformity}
\newacronym{CMOS}{CMOS}{complementary metal–oxide–semiconductor}
\newacronym{TRL}{TRL}{technology readiness level}

\newacronym{FOV}{FOV}{field-of-view}
\newacronym{NIR}{NIR}{near-infrared}
\newacronym{PV}{PV}{Peak-to-Valley}
\newacronym{MRF}{MRF}{Magnetorheological finishing}
\newacronym{AO}{AO}{Adaptive Optics}
\newacronym{TTP}{TTP}{tip, tilt, and piston}
\newacronym{FPS}{FPS}{fine pointing system}
\newacronym{SHWFS}{SHWFS}{Shack-Hartmann Wavefront Sensor}
\newacronym{OAP}{OAP}{off-axis parabola}
\newacronym{LGS}{LGS}{laser guide star}
\newacronym{WFCS}{WFCS}{wavefront control system}
\newacronym{OPD}{OPD}{optical path difference}

\newacronym{acs}{ACS}{Attitude Control System}
\newacronym{orsa}{ORSA}{Ogive Recovery System Assembly}
\newacronym{gse}{GSE}{Ground Station Equipment}
\newacronym{FSM}{FSM}{Fast Steering Mirror}

\newacronym{WFS}{WFS}{wavefront sensor}
\newacronym{LSI}{LSI}{Lateral Shearing Interferometer}
\newacronym{VVC}{VVC}{Vector Vortex Coronagraph}
\newacronym{VNC}{VNC}{Visible Nulling Coronagraph}
\newacronym{CGI}{CGI}{Coronagraph Instrument}
\newacronym{IWA}{IWA}{Inner Working Angle}
\newacronym{OWA}{OWA}{Outer Working Angle}
\newacronym{NPZT}{N-PZT}{Nuller piezoelectric transducer}
\newacronym{ZWFS}{ZWFS}{Zernike wavefront sensor}
\newacronym{SPC}{SPC}{Shaped Pupil Coronagraph}
\newacronym{HLC}{HLC}{Hybrid-Lyot Coronagraph}
\newacronym{ADI}{ADI}{angular differential imaging}
\newacronym{RDI}{RDI}{reference differential imaging}

\newacronym{HST}{HST}{Hubble Space Telescope}
\newacronym{GPS}{GPS}{Global Positioning System}
\newacronym{ISS}{ISS}{International Space Station}
\newacronym[description=Advanced CCD Imaging Spectrometer]{acis}{ACIS}{Advanced \gls{ccd} Imaging Spectrometer}
\newacronym{stis}{STIS}{\textit{Space Telescope Imaging Spectrograph}}
\newacronym{mcp}{MCP}{Microchannel Plate}
\newacronym{jwst}{JWST}{$\textit{James Webb Space Telescope}$}
\newacronym{fuse}{FUSE}{$\textit{FUSE}$}
\newacronym{galex}{GALEX}{$\textit{Galaxy Evolution Explorer}$}
\newacronym{spitzer}{Spitzer}{$\textit{Spitzer Space Telescope}$}
\newacronym{mips}{MIPS}{Multiband Imaging Photometer for \gls{spitzer}}
\newacronym{gissmo}{GISSMO}{Gas Ionization Solar Spectral Monitor}
\newacronym{iue}{IUE}{International Ultraviolet Explorer}
\newacronym{spinr}{SPINR}{$\textit{Spectrograph for Photometric Imaging with Numeric Reconstruction}$}
\newacronym{imager}{IMAGER}{$\textit{Interstellar Medium Absorption Gradient Experiment Rocket}$}
\newacronym{TPF-C}{TPF-C}{Terrestrial Planet Finder Coronagraph}
\newacronym{RAIDS}{RAIDS}{Atmospheric and Ionospheric Detection System }
\newacronym{mama}{MAMA}{Multi-Anode Microchannel Array}
\newacronym{ATLAST}{ATLAST}{Advanced Technology Large Aperture Space Telescope}
\newacronym{PICTURE}{PICTURE}{Planet Imaging Concept Testbed Using a Rocket Experiment}
\newacronym{LITES}{LITES}{Limb-imaging Ionospheric and Thermospheric
Extreme-ultraviolet Spectrograph}
\newacronym{LBT}{LBT}{Large Binocular Telescope}
\newacronym{LBTI}{LBTI}{Large Binocular Telescope Interferometer}
\newacronym{KIN}{KIN}{Keck Interferometer Nuller}
\newacronym{SHARPI}{SHARPI}{Solar High-Angular Resolution Photometric Imager}
\newacronym{IRAS}{IRAS}{Infrared Astronomical Satellite}
\newacronym{HARPS}{HARPS}{High Accuracy Radial velocity Planetary}
\newacronym{hstSTIS}{STIS}{Space Telescope Imaging Spectrograph}
\newacronym{spitzerIRAC}{IRAC}{Infrared Array Camera}
\newacronym{spitzerMIPS}{MIPS}{Multiband Imaging Photometer for Spitzer}
\newacronym{spitzerIRS}{IRS}{Infrared Spectrograph}
\newacronym{CHARA}{CHARA}{Center for High Angular Resolution Astronomy}
\newacronym{wfirst-afta}{WFIRST-AFTA}{Wide-Field InfrarRed Survey
Telescope-Astrophysics Focused Telescope Assets}
\newacronym{GPI}{GPI}{Gemini Planet Imager}
\newacronym{WFIRST}{WFIRST}{Wide-Field InfrarRed Survey Telescope}
\newacronym{HabEx}{HabEx}{Habitable Exoplanet Observatory Mission Concept}
\newacronym{LUVOIR}{LUVOIR}{Large UV/Optical/Infrared Surveyor}
\newacronym{FGS}{FGS}{Fine Guidance Sensor}
\newacronym{STIS}{STIS}{Space Telescope Imaging Spectrograph}
\newacronym{MGHPCC}{MGHPCC}{Massachusetts Green High Performance
Computing Center}
\newacronym{WISE}{WISE}{Wide-field Infrared Survey Explorer}
\newacronym{ALMA}{ALMA}{Atacama Large Millimeter Array}
\newacronym{GRAIL}{GRAIL}{Gravity Recovery and Interior Laboratory}

\newacronym{AURIC}{AURIC}{The Atmospheric Ultraviolet Radiance Integrated Code} 
\newacronym{FFT}{FFT}{Fast Fourier Transform  }
\newacronym{MODTRAN}{MODTRAN   }{ MODerate resolution atmospheric TRANsmission }
\newacronym{idl}{IDL}{$\textit {Interactive Data Language}$}
\newacronym[sort=NED,description=NASA/IPAC Extragalactic Database]{ned}{NED}{\gls{nasa}/\gls{ipac} Extragalactic Database}
\newacronym{iraf}{IRAF}{Image Reduction and Analysis Facility}
\newacronym{wcs}{WCS}{World Coordinate System}
\newacronym{pegase}{PEGASE}{$\textit{Projet d'Etude des GAlaxies par Synthese Evolutive}$}
\newacronym{dirty}{DIRTY}{$\textit{DustI Radiative Transfer, Yeah!}$}
\newacronym{CUDA}{CUDA}{Compute Unified Device Architecture}

\newacronym{MSIS}{MSIS}{Mass Spectrometer Incoherent Scatter Radar}
\newacronym{nmf2}{$N_m$}{F2-Region Peak density}
\newacronym{hmf2}{$h_m$}{F2-Region Peak height}
\newacronym{H}{$H$}{F2-Region Scale Height}

\newacronym{isr}{ISR}{Incoherent Scatter Radar}
\newacronym[description=TLA Within Another Acronym]{twaa}{TWAA}{\gls{tla} Within Another Acronym}
\newacronym[plural=SNe, firstplural=Supernovae (SNe)]{sn}{SN}{Supernova}
\newacronym{EUV}{EUV}{Extreme-Ultraviolet }
\newacronym{EUVS}{EUVS}{\gls{EUV} Spectrograph}
\newacronym{F2}{F2}{Ionospheric Chapman F Layer }
\newacronym{F10.7}{F10.7}{ 10.7 cm radio flux [10$^{-22}$ W m$^{-2}$ Hz$^{-1}$]  }
\newacronym{FUV}{FUV}{ Far-Ultraviolet }
\newacronym{IR}{IR}{Infrared}
\newacronym{MUV}{MUV}{Mid-Ultraviolet }
\newacronym{NUV}{NUV}{Near-Ultraviolet }
\newacronym{O$^+$}{O$^+$}{Singly Ionized Oxygen  Atom }
\newacronym{OI}{OI}{Neutral Atomic Oxygen Spectroscopic State }
\newacronym{OII}{OII}{Singly Ionized Atomic Oxygen Spectroscopic State }
\newacronym{PSF}{PSF}{Point Spread Function}
\newacronym{$R_E$}{$R_E$}{ Earth Radii [$\approx$ 6400 km]  }
\newacronym{RV}{RV}{Radial Velocity}
\newacronym{UV}{UV}{Ultraviolet }
\newacronym{WFE}{WFE}{Wavefront Error}
\newacronym{sed}{SED}{Spectral Energy Distribution}
\newacronym{nir}{NIR}{near-infrared}
\newacronym{mir}{MIR}{mid-infrared}
\newacronym{ir}{IR}{infrared}
\newacronym{uv}{UV}{ultraviolet}
\newacronym[plural=PAHs, firstplural=Polycyclic Aromatic Hydrocarbons (PAHs)]{pah}{PAH}{Polycyclic Aromatic Hydrocarbon}
\newacronym{obsid}{OBSID}{Observation Identification}
\newacronym{SZA}{SZA}{Solar Zenith Angle}
\newacronym{TLE}{TLE}{Two Line Element set}
\newacronym{DOF}{DOF}{degrees-of-freedom}
\newacronym{PZT}{PZT}{lead zirconate titanate}
\newacronym{ADCS}{ADCS}{attitude determination and control system}
\newacronym{COTS}{COTS}{commercial off-the-shelf}
\newacronym{CDH}{C$\&$DH}{command and data handling}
\newacronym{EPS}{EPS}{electrical power system}

\newacronym{PCA}{PCA}{Principal Component Analysis}
\newacronym{fwhm}{FWHM}{Full-Width-Half Maximum}
\newacronym{RMS}{RMS}{root mean squared}
\newacronym{RMSE}{RMSE}{root mean squared error}
\newacronym{MCMC}{MCMC}{Marcov chain Monte Carlo}
\newacronym{DIT}{DIT}{Discrete Inverse Theory}
\newacronym{SNR}{SNR}{signal-to-noise ratio}
\newacronym{PSD}{PSD}{power spectral density}

\let\oldadded\added
\renewcommand{\added}[1]{\oldadded{{#1}}}

\newglossarystyle{symbunitlong}{%
\setglossarystyle{long3col}
  {\end{longtable}}%

}
\glsdisablehyper 

\newglossary[slg]{notation}{syi}{syg}{Symbolslist} 

\makenoidxglossaries

\newglossaryentry{alpha}
{%
  type=notation,
  symbol={$\alpha$},
  description={},
  name={power law index of the disturbance PSD},
  sort={S}%
} 

\newglossaryentry{beta_p}
{%
  type=notation,
  name={wavefront error PSD normalization constant},%
  symbol={$\beta_{OPD}$},%
  description={},%
  sort={S}%
} 

\newglossaryentry{ETFf}
{%
  type=notation,
  symbol={$ETF(f)$},
  description={},
  name={ Error Transfer Function},
  sort={S}%
} 

\newglossaryentry{f}
{%
  type=notation,
  symbol={$f$},
  description={[Hz]},
  name={disturbance frequency},
  sort={S}%
} 
\newglossaryentry{F_gamma}
{%
  type=notation,
  name={ the photon rate}, 
  symbol={$F_\gamma$},
  description={[photons/second] },
  sort={S}%
} 
\newglossaryentry{f_s}
{%
  type=notation,
  symbol={$f_s$},
  description={[Hz]},
  name={sampling frequency},
  sort={S}%
} 

\newglossaryentry{lambda_wfs}
{%
  type=notation,
  name={wavefront sensing wavelength},
  symbol={$\lambda_{WFS}$},
  description={  [nm] },
  sort={S}%
} 
\newglossaryentry{MFD}
{%
  type=notation,
  symbol={$MFD$},
  description={[um]},
  name={mode-field diameter},
  sort={S}
}
\newglossaryentry{NTFf}
{%
  type=notation,
  symbol={$NTF(f)$},
  description={},
  name={ Noise Transfer Function},
  sort={S}%
}
\newglossaryentry{sigma_10}
{%
  type=notation,
  name={1$\sigma$ stability over 10 minutes},
  symbol={$\sigma_{10}$},
  description={[pm]},
  sort={S}%
} 
\newglossaryentry{sigma_ron}
{%
  type=notation,
  name={detector readout noise },
  symbol={$\sigma_{ron}$},
  description={[electrons] },
  sort={S}%
} 
\newglossaryentry{sigma_WFE}
{%
  type=notation,
  name={$\sigma_{WFE}$},
  description={\Gls{WFE} \gls{RMSE} [nm]},
  sort={S}%
} 
\newglossaryentry{S_0}
{%
  type=notation,
  symbol={$S_0$},
  description={electrons per pixel},
  name={Average entrance pupil intensity},
  sort={S}%
} 
\newglossaryentry{tau_wfs}
{%
  type=notation,
  name={wavefront sensor exposure time},
  symbol={$\tau_{wfs}$},
  description={ [sec] },
  sort={S}%
} 
\newglossaryentry{T_0}
{%
  type=notation,
  symbol={$T_0$},
  description={minutes},
  name={inverse of the disturbance PSD knee frequency},
  sort={S}%
} 
\newglossaryentry{theta}
{%
  type=notation,
  symbol={$\theta$},
  description={radian},
  name={transmitter half-angle divergence},
  sort={S}%
} 
\newglossaryentry{w_0}
{%
  type=notation,
  symbol={$w_0$},
  description={[m]},
  name={beam waist},
  sort={S}
}
\newglossaryentry{zeta_c}
{%
  type=notation,
  symbol={$\zeta_c$},
  description={},
  name={planet-star flux ratio},
  sort={S}%
} 

\newglossaryentry{x_0}
{%
  type=notation,
  symbol={$x_0$},
  description={[radian]},
  name={pointing error},
  sort={S}%
} 
\newglossaryentry{x}
{%
  type=notation,
  symbol={$x$},
  description={[radian]},
  name={radial displacement from gaussian beam},
  sort={S}%
} 
\newglossaryentry{w}
{%
  type=notation,
  symbol={$w$},
  description={[radian]},
  name={gaussian beam width},
  sort={S}%
} 
\newglossaryentry{phi}
{%
  type=notation,
  symbol={$\phi$},
  description={[radian]},
  name={wavefront sensor phase error},
  sort={S}%
} 
\newglossaryentry{Tp}
{%
  type=notation,
  symbol={$T_p$},
  description={ratio},
  name={System Throughput and QE},
  sort={S}%
}  
\newacronym{RECONS}{RECONS}{Research Consortium On Nearby Stars}

\shorttitle{Laser Guide Star for Large Segmented-Aperture Space Telescopes, Part I}
\shortauthors{Douglas, E.S. et al.}

\begin{document}

\title{Laser Guide Star for Large Segmented-Aperture Space Telescopes, Part I: Implications for Terrestrial Exoplanet Detection and Observatory Stability}

\author{E.S. Douglas}
\affiliation{ Department of Aeronautics and Astronautics, Massachusetts Institute of Technology}
\author{J. R. Males}
\affiliation{Steward Observatory, University of Arizona}
\author{J. Clark}
\affiliation{ Department of Aeronautics and Astronautics, Massachusetts Institute of Technology}
\author{O. Guyon}
\affiliation{Steward Observatory, University of Arizona}
\author{J. Lumbres} 
\affiliation{Steward Observatory, University of Arizona}
\author{W. Marlow} 
\affiliation{ Department of Aeronautics and Astronautics, Massachusetts Institute of Technology}
\author{K.L. Cahoy}
\affiliation{ Department of Aeronautics and Astronautics, Massachusetts Institute of Technology}
\affiliation{ Department of Earth, Atmospheric, and Planetary Science, Massachusetts Institute of Technology}

\begin{abstract}
Precision wavefront control on future segmented-aperture space telescopes presents significant challenges, particularly in the context of high-contrast exoplanet direct imaging. We present a new wavefront control architecture that translates the ground-based artificial guide star concept to space with a laser source aboard a second spacecraft, formation flying within the telescope field-of-view.
We  describe the motivating problem of mirror segment motion and develop wavefront sensing requirements as a function of guide star magnitude and segment motion power spectrum. 
Several sample cases with different values for transmitter power, pointing jitter, and wavelength are presented to illustrate the  advantages and challenges of having a non-stellar-magnitude noise limited wavefront sensor for space telescopes. 
These notional designs allow increased control authority, potentially relaxing  spacecraft stability requirements by two orders of magnitude, and increasing terrestrial exoplanet discovery space by allowing high-contrast observations of stars of arbitrary brightness.
\end{abstract}
\newpage
\clearpage

\newpage 
\printnoidxglossary[type=notation,style=symbunitlong,title={Table of Symbols},nonumberlist]

 \section{Introduction}\label{sec:intro}
Reflected light imaging of terrestrial exoplanets with space telescopes requires both large apertures and extreme instrument stability.
The brightest observed flux ratio (\glssymbol{zeta_c}) between a planet with an Earth-like albedo and radius  and a Sun-like host star is approximately $10^{-10}$ \added{or 25 magnitudes}, with deeper contrasts at intermediate phases and spectral absorption features \citep{woolf_spectrum_2002,turnbull_spectrum_2006,robinson_earth_2011}.
As an alternative to coronagraphs, formation flying large ($>40$ m) external occulters, or starshades, provide high sensitivity  in exchange for  long wait times between targets to reposition the occulter.  
Given this high overhead, starshades may be preferable for spectroscopy of known exoplanets, while coronagraphs may provide higher yields in blind searches \cite{stark_maximized_2016}
In order to discover and/or characterize a significant number of Earth-like planets in a survey of nearby stars within a typical five year mission lifetime, apertures greater than 4 m diameter and coronagraphic attenuation of starlight (i.e. contrast) to below $10^{-11}$ are likely needed \citep{stark_maximizing_2014,stark_direct_2016}. 

\cite{stark_lower_2015} modeled detection limits for habitable-zone Earth-like exoplanets with a 10 meter space observatory for a total mission exposure time of 1 year (including spectral characterization).
By holding other model assumptions constant, they found a \deleted{weak} power law dependence of yield on contrast of $\zeta_C^{-0.1}$. 
In the \cite{stark_lower_2015} example case, decreasing the contrast from $10^{-10}$ to $10^{-9}$ decreases the mission yield from 26 to 14 Earth-like planets, underscoring the importance of maximizing contrast.

Internal coronagraphic instruments which attenuate starlight and allow exoplanet detection at small separations are highly sensitive to wavefront errors (c.f. \cite{serabyn_nulling_2000,traub_direct_2010}).
The wavefront must be sufficiently stable in order to sense, control, and  subtract systematic leakage (commonly known as ``speckles'', c.f. \cite{racine_speckle_1999,perrin_structure_2003}). 
In order to maximize collecting area and resolution, large apertures (4 m - 15 m) are also needed.
To achieve such large apertures, missions such as the proposed \gls{LUVOIR} concept, are expected to use primary mirrors made up of multiple meter-scale  segments \citep{eisenhower_atlast_2015}. 

Segment motion is a mid-spatial frequency wavefront error which causes speckles inside a coronagraph dark hole \citep{ruane_performance_2017, leboulleux_pair-based_2018}.
Thus, the wavefront error in the segment tip, tilt, and piston modes must be highly stabilized for imaging and spectroscopy of Earth-like exoplanets in visible light.
A variety of efforts are underway to develop and test coronagraphs for segmented apertures \citep{miller_ua_2015,ndiaye_apodized_2016,ruane_performance_2017,hicks_segmented_2018, martinez_segmented_2018}.
\Gls{RMS} \gls{WFE} stabilities below 10 picometers are commonly  specified to reach the required flux ratios \citep{lyon_space_2012,bolcar_large_2017}.
As discussed in Section \ref{sec:closed_loop}, the particular \gls{WFE} requirements depend on the temporal  \gls{PSD} of the segment motion.

Observatories on the ground have demonstrated alignment of telescopes  made up of multiple segments. 
For example, the Multiple Mirror Telescope \citep{beckers_performance_1982} alignment is achieved by actively controlling a segmented  secondary mirror, while for the W.M. Keck telescope, alignment is achieved by controlling primary segment position \citep{jared_w._1990}. 
Similar systems are planned for nanometer level control of upcoming thirty-meter-class telescope segments \citep{macintosh_extreme_2006-1,gonte_active_2008,troy_conceptual_2008,bouchez_giant_2012}.

Different means of sensing segment motion to picometer levels have been proposed: edge sensors, wavefront sensing using the target star, or internal metrology \citep{feinberg_ultra-stable_2017}.
Wavefront sensing using target starlight minimizes calibration errors between sensors and the science image; however, photon noise limits  wavefront sensing (and science observations)  to bright nearby stars \citep{lyon_space_2012,stahl_engineering_2013,stahl_preliminary_2015}. 
Contrast depends on wavefront sensing and control, which in turn requires sufficient flux for effective wavefront sensing. 
A guide star of arbitrary brightness  offers the potential to significantly increase the yield of a survey, by increasing the sensitivity of a given observatory to exoplanets even for dim targets and can significantly increase yield for large aperture space telescopes.

We present a new approach to wavefront sensing, employing a bright formation flying calibration source, serving as an artificial guide star and enabling high-cadence segment control during  coronagraph observations of stellar systems regardless of host star magnitude.
 \deleted{Additionally, this approach also could permit relaxed telescope stability requirements,  decreasing  mass and cost.}

Artificial guide stars were developed for ground-based astronomical telescopes several decades ago. \cite{foy_feasibility_1985} proposed using laser light from the ground to illuminate a bright artificial star at high  altitudes as a reference for ground-based adaptive optics systems. 
This was soon demonstrated by exciting mesospheric sodium \citep{thompson_experiments_1987}, an approach that become a widely used means of improving adaptive optics performance (e.g. \cite{max_image_1997,wizinowich_w._2006,holzlohner_optimization_2010}).  

Several authors have considered a space-borne \gls{LGS} for use with ground-based telescopes, which would provide a diffraction-limited point source and operate at lower power than atmospheric backscatter guide stars.
 \cite{greenaway_pharos_1990} proposed a laser source in cis-lunar orbit for observing low-declination astronomical targets.
Similarly, drones have been proposed as platforms for downward looking laser guide stars \citep{basden_artificial_2018}.
 \cite{marlow_laser-guide-star_2017} proposed a CubeSat nanosatellite in geosynchronous orbit for astronomical imaging and space situational awareness from the ground. 
 
 Adaptive optics systems for use with ground-based telescopes primarily mitigate aberrations caused by atmospheric turbulence. This is different than the motivation for a space-based laser guide star paired with a large aperture segmented space telescope. Instead of atmospheric turbulence, a space-based laser guide star enables correction of static and dynamic wavefront errors caused by onboard structural, thermal, and optical sources.
Building on  the concept of a CubeSat  \gls{LGS} described by \cite{marlow_laser-guide-star_2017}, this work explores the adaptation of a small  \gls{LGS} spacecraft to enable precise wavefront sensing of a large segmented-aperture space telescope.
While spacecraft formation flight  is challenging,  two-spacecraft precision formation flight without use of the Global Positioning System for navigation has been demonstrated by  the Gravity Recovery and Interior Laboratory mission in lunar orbit with $\sim1 \mu$m/s accuracy \citep{smith_high-accuracy_2016} and it may be a viable solution for missions of the size and complexity of \gls{LUVOIR}. 
Section \ref{sec:methods}  describes basic parameters of a laser guide star spacecraft, the problem of segment motion in the context of high-contrast imaging, and develops a notional \gls{WFS}.
Section \ref{sec:results}   presents a laser guide start mission architecture to meet the performance requirements of a large segmented aperture telescope mission.
Section \ref{sec:discussion} discusses the science impact of maintaining contrast while observing dim target stars and the engineering impact of relaxing telescope stability requirements.
Section \ref{sec:summary} provides a summary of the benefits of a laser guide star system, and discusses ongoing experiments and next steps toward developing a laser guide star technology demonstration.

\section{ Methods: Establishing Telescope Stability Requirements for Earth-like Planet Detection}\label{sec:methods}
Large aperture space telescope designs call for deployed segmented apertures in order to fit within launch vehicle fairings (e.g. \cite{postman_advanced_2009,stahl_luvoir_2016}). 
A low-mass segmented telescope is easier to launch, package, and maneuver;   while a stiffer, more massive telescope is easier to align and stabilize against disturbances.
Segment motion arises primarily from vibrations in the spacecraft, due to imperfections in reaction wheel bearings and balance,  or  variations in thruster performance \citep{mier-hicks_electrospray_2017} which are transmitted by the relatively flexible, low-mass spacecraft structure \citep{bronowicki_vibration_2006}.
\cite{stahl_preliminary_2015} finds a 0.25 minute wavefront sensing cadence is required for a 12 m telescope observing a $m_\textrm{V}$=5 star, with a stability ten times longer ($\gtrsim2.5$ minutes) to ensure control system performance (see also \cite{lyon_space_2012,stahl_engineering_2013}).
Depending on the coronagraph, the wavefront error requirement can be significantly relaxed for lower spatial order modes, such as global tilt or focus \citep{ruane_performance_2017}.
However, segment motion primarily contributes at higher spatial frequencies, degrading contrast at planet-star separations of significant interest.

\subsection{Impact of Wavefront Error on Coronagraph Contrast}
Modeling of exoplanet yield versus contrast (e.g. \cite{stark_lower_2015}) typically depends on a constant contrast floor from the \gls{IWA} to the \gls{OWA}. 
However, the sensitivity of  coronagraphs varies as a function of radius from the star. 
In order to estimate the sensitivity of a coronagraphic telescope to exoplanets, we define contrast as the raw instrumental  ratio of spurious speckle light to the peak of the stellar  \gls{PSF}. 

The speckle brightness depends on the sum of the amplitudes of system wavefront errors at a particular spatial frequency  \cite[Equation 123]{traub_direct_2010}.

\begin{figure}
    \centering
            \plottwo{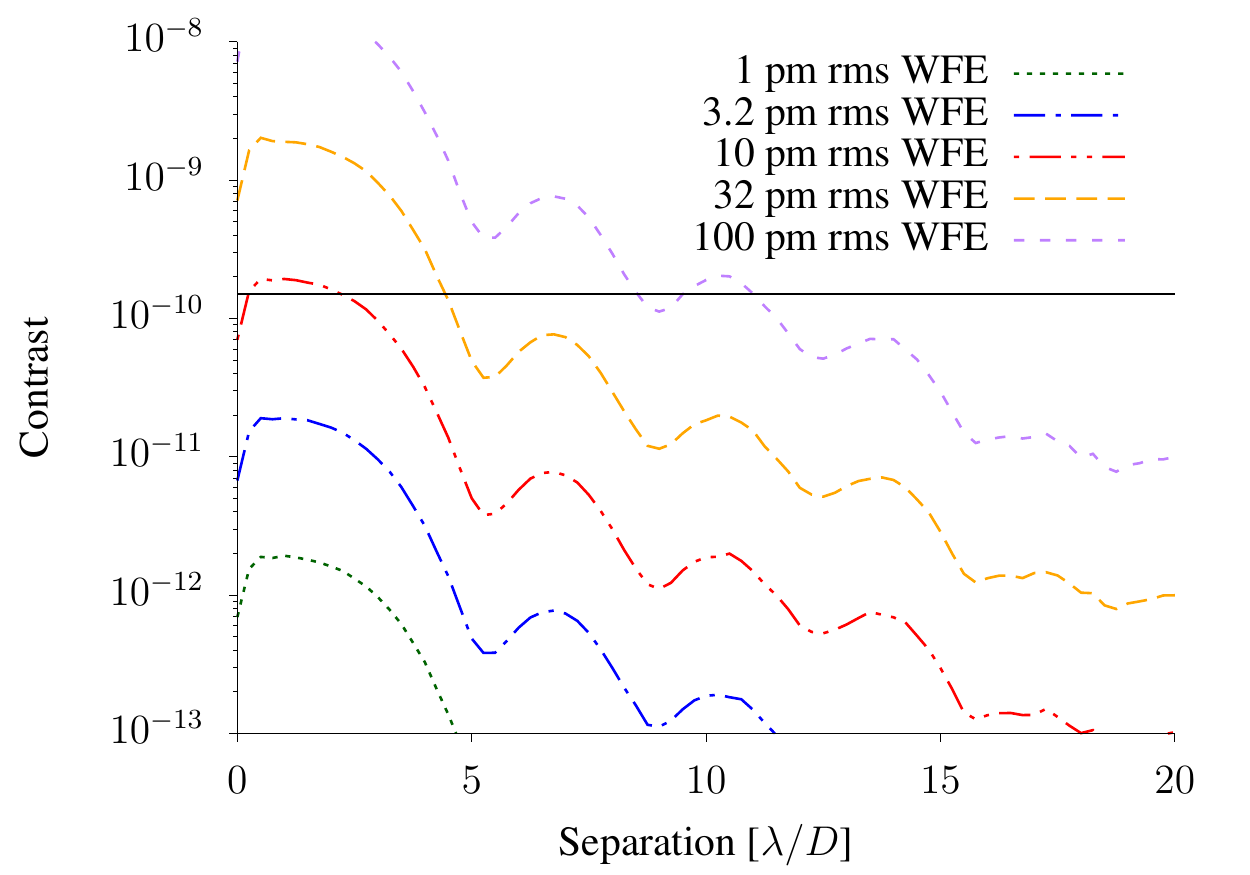}{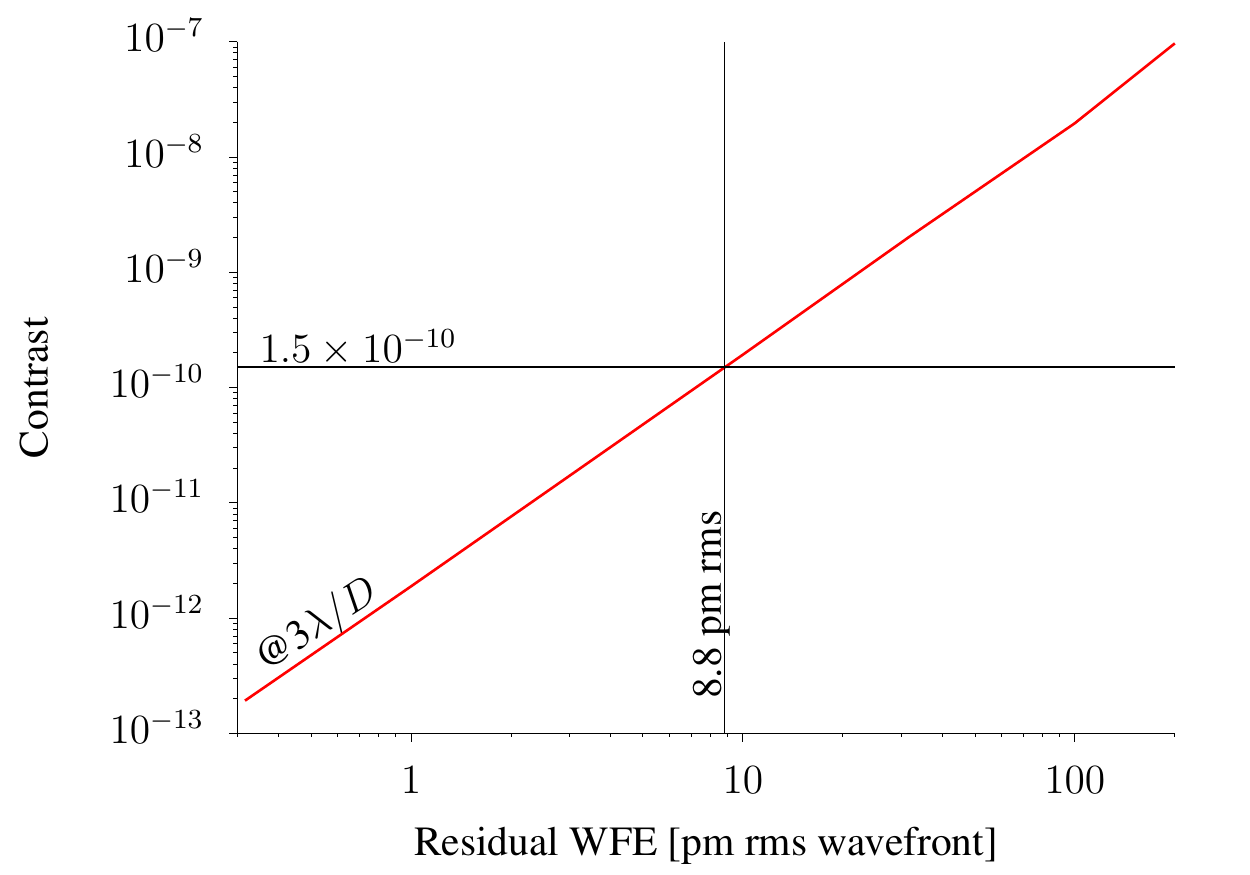}
        \caption{Left: Numerical simulation of contrast versus \gls{RMS} segment motion wavefront error for an ideal coronagraph. 
        Curves show decreasing wavefront error from top to bottom. 
        A horizontal line indicates the contrast of an Earth-radius exoplanet with an albedo of 0.25 at quadrature. 
        Right: Raw contrast versus residual \gls{WFE} at 3 $\lambda/D$, showing that 10 pm is a good approximation for a \gls{WFE} requirement for detection of Earth-like planets with a segmented-aperture telescope.} 
        \label{fig:contrast_vs_disturbance}
    \end{figure}

In order to assess the influence of segment motion on speckle brightness we developed a numerical model in the Fraunhofer diffraction regime of an ideal coronagraph  \citep{males_ground-based_2018} and a segmented primary mirror. 
Contrast curves are shown in the left panel of  Fig.~\ref{fig:contrast_vs_disturbance} for a variety of  \gls{RMS} segment wavefront disturbances and compared to a Earth-like exoplanet contrast (horizontal line).
 The contrast as a function of angle peaks near  1$\lambda/D$, where the low-spatial-frequency segment motion has the largest impact; 10 pm \gls{RMS} (dot-dot-dash line) corresponds to a contrast of approximately 1.5$\times 10^{-10}$.
 Earth-like planet yield is highly sensitive to a coronagraph's \gls{IWA} \citep{stark_lower_2015} and these low-spatial frequency errors likewise strongly impact sensitivity.
These results are generally consistent with recent work on segmented mirror impact on coronagraph contrast (e.g. \cite{ruane_performance_2017,leboulleux_pair-based_2018}), with the caveat that here the segment tip-tilt and piston modes have been normalized to contribute equal \gls{RMS} \gls{OPD} disturbances.

\subsection{Wavefront Error Simulation}
\begin{figure}
\centering
\includegraphics{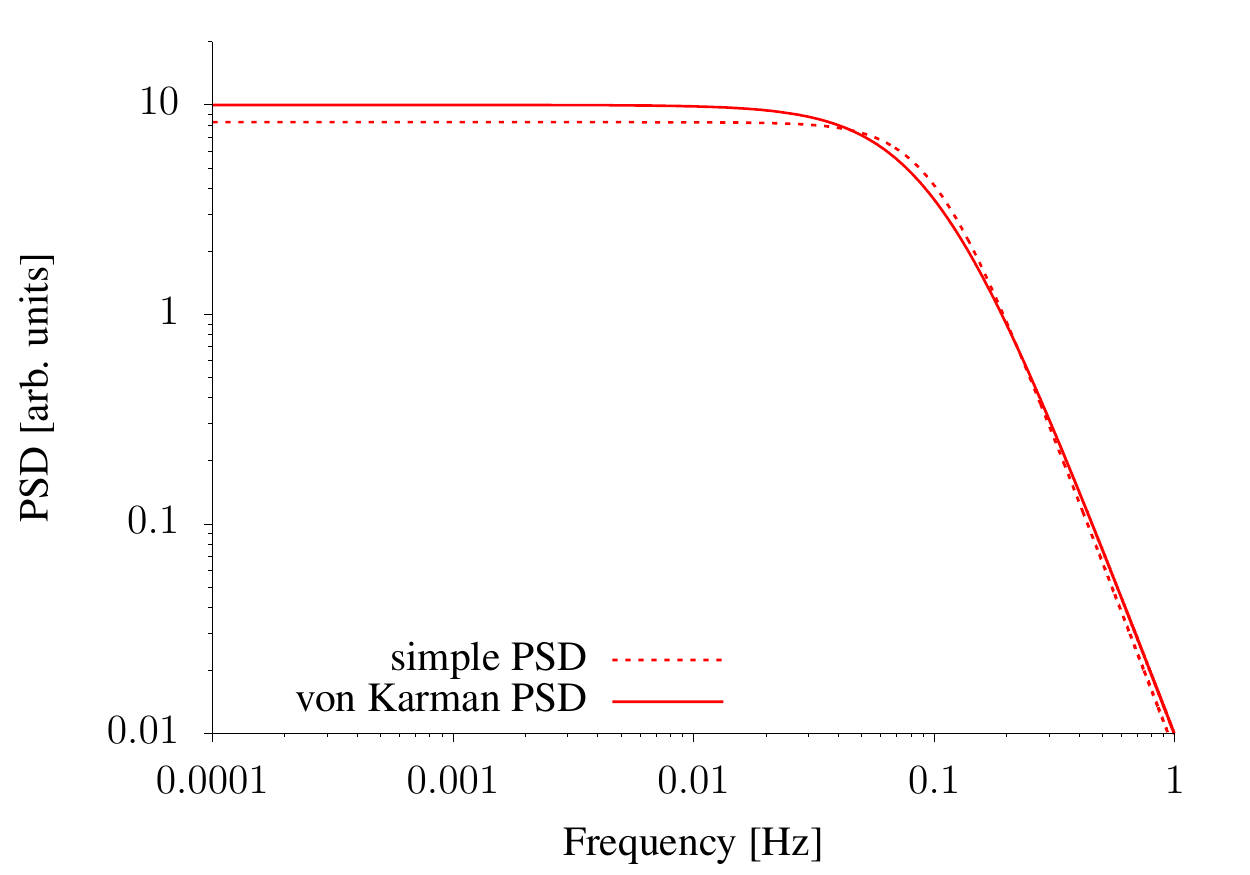}
\caption{Comparison   with arbitrary units of the simple PSD form of Equation \ref{eqn:lyon_psd} with the more general von Karm\`an or K-correlation form of Equation \ref{eqn:vk_psd}. 
For equal knee frequencies, $f_n=$0.1 Hz in this case, the forms are very similar. 
 We adopt the von Karm\`an PSD in this work. \label{fig:vkPSDcomp}}
\end{figure}

This section will explore the relationship between mechanical stability and incident photon rate by applying a control law to  the \gls{PSD} defining segment motion. 
This will lay the groundwork for setting design constraints on artificial laser guide stars.
A realized optomechanical system will have time-dependent \glspl{OPD} arising from a variety of mechanical disturbances (c.f. \cite{bronowicki_vibration_2006,shi_low_2016}).   
To constrain the problem, we assume a smooth \gls{PSD}.
The form of the functional \gls{PSD} we have chosen for modeling the longer timescale motion of primary mirror segments is similar to previous work, but with a few key differences. 
Previous work by \cite{2012OptEn..51a1002L}  assumed a \gls{OPD} PSD with respect to  frequency, \glssymbol{f}, of the form:
\begin{equation}
\mathcal{T}(f) \propto \frac{1}{1 + \left(\frac{f}{f_n}\right)^\alpha}.
\label{eqn:lyon_psd}
\end{equation}
Here \glssymbol{alpha} is a power law constant and $f_n$ is the ``knee frequency'' where the distribution rolls off.
A form commonly used to model optical surfaces \citep{1986SPIE..645..107C,1991SPIE.1530...71C,1996SPIE.2775..240T,2009SPIE.7426E..0IH} is the K-correlation model, which in optical turbulence modeling is known as the von Karm\`an PSD \citep{1998aoat.book.....H,2005OptProptSPIE}. We adopt the following form as the PSD of the optical path difference due to segment motion:
\begin{equation}
\mathcal{T}_{OPD}(f)=\frac{ \beta_{OPD}^2}{\left(f_o^2 + f^2\right)^{\alpha/2}}\label{eqn:vk_psd}
\end{equation}
Here  $\beta_{OPD}$ is a normalization constant and $f_o$ is the knee frequency, which is defined in terms of an ``outer time'' $T_0$ by $f_0=1/T_0$ (in analogy with the outer scale in turbulence). 
 This has a slightly different form from that used by \cite{2012OptEn..51a1002L} for both spatial and temporal PSDs. 
    Fig.~\ref{fig:vkPSDcomp} directly compares the two similar forms. $f_0$ is essentially equivalent to their ``drift frequency'' $f_n$. 
  In addition to its more general use in the literature, we prefer the \gls{PSD} in Equation \ref{eqn:vk_psd} to that Equation \ref{eqn:lyon_psd} due to its simpler behavior as $f_0\rightarrow 0$, where it trivially becomes pure $1/f^\alpha$ noise. 

Whether or not such PSDs are integrable depends on $f_0$ and $\alpha$.  In order to allow any value of these parameters, we adopt a band-limited stability specification.  We call this \glssymbol{sigma_10}, or the ``\gls{RMS} in 10 min'', i.e. ``10 pm \gls{RMS} in 600 sec''. 
 We normalize the PSD accordingly, from the frequency corresponding to 10 minutes to one half of \glssymbol{f_s}, the sampling frequency of the wavefront control system, i.e.
\begin{equation}
\beta_{OPD}^2 = \frac{\sigma_{10}^2} {\displaystyle\int_{\frac{1}{600 \textrm{ sec}}}^{f_s/2} \left(f_o^2 + f^2\right)^{-\alpha/2} df}.
\end{equation}
The PSD of measurement noise is given by \cite{2018JATIS...4a9001M} as:
\begin{equation}
\mathcal{T}_{p}(f) = \frac{\beta_p^2}{F_\gamma \tau_{wfs} + n_{pix}\sigma_{ron}^2}\left(\frac{\tau_{wfs}}{2}\right),\label{eq:tpf}
\end{equation}
where we are ignoring background noise sources which will not significantly impact the shape of the \gls{PSD}.
 $n_{pix}$ is the number of detector pixels used, each with readout noise \glssymbol{sigma_ron}.  See Table \ref{tab:control_params} for assumed noise values and the number of pixels per segment.
 \added{Since \glssymbol{tau_wfs} appears in both the numerator and denominator, when \glssymbol{sigma_ron} is set to zero the  \glssymbol{tau_wfs} cancels and for Equation \ref{eq:tpf} } the measurement noise \added{PSD} does not depend on wavefront sensor exposure time, \glssymbol{tau_wfs}, given a noiseless (or low-noise) detector\added{\footnote{Note that the total measurement noise depends on the integral of the \gls{PSD} and does depend on the exposure time.}}.
 
 \added{The observed \gls{PSD}  depends on the measured system losses. For the current \gls{LUVOIR} design, a system optical throughput to the wavefront sensor of 37.39\% at 532 nm, and  41.20\% at 980 nm is expected \citep{bolcar_luvoir_2018}.
 We set a conservatively low throughput of 10\%  to account for reflective losses, as well as other terms decreasing the photon count rate, including the detector quantum efficiency, surface contamination and degradation,  and aperture obscuration. }

\begin{figure}
\centering
\includegraphics[width=3in]{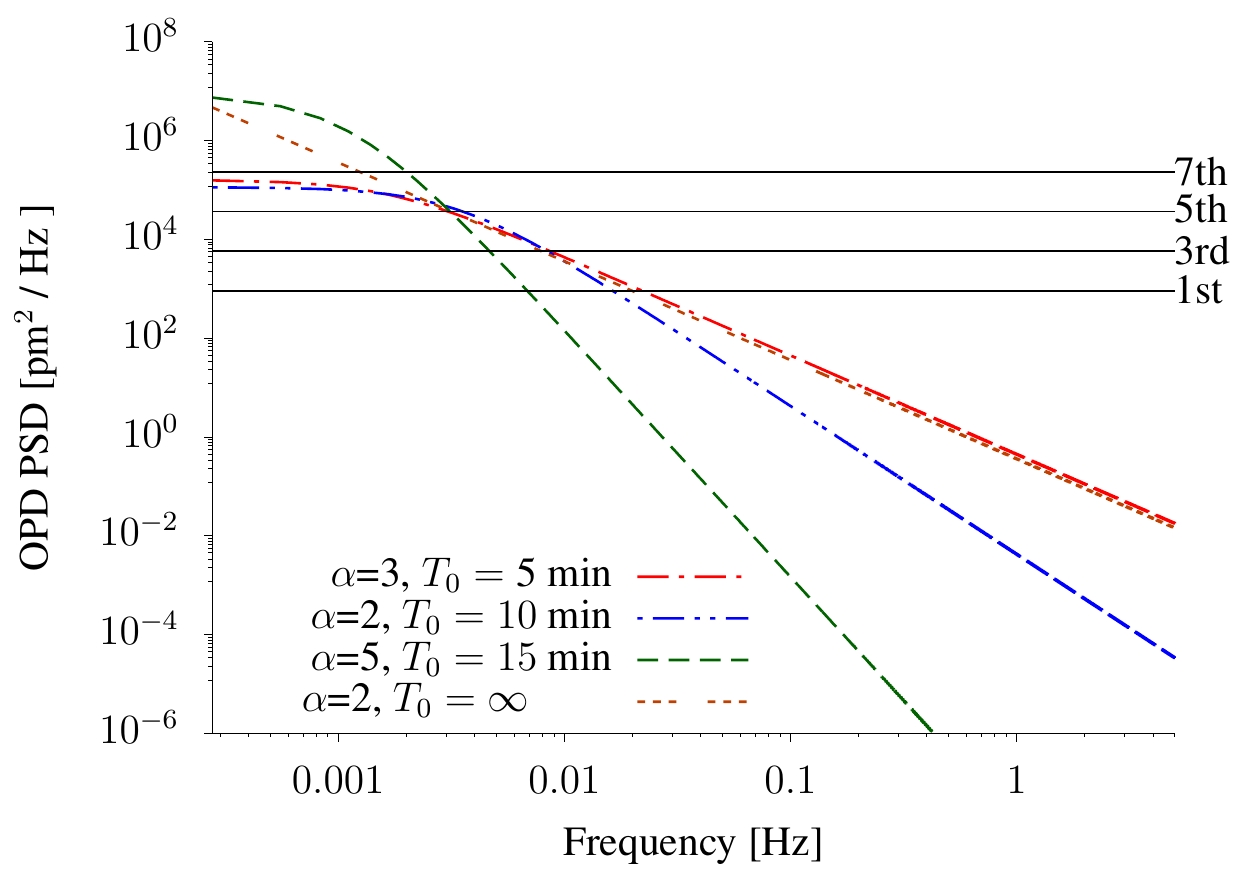}
\caption{Comparison of PSDs with with different power law constants ($\alpha$) and different $T_0$ outer times. The dashed lines show the optical path difference PSDs and the horizontal solid lines show the measurement noise floor for stars of different magnitudes.  
For $m_\textrm{V}$=7 and dimmer stars measurement noise dominates.
Even for a steep $\alpha$=5 \gls{PSD}, stellar noise will dominate above 0.01 Hz for an $m_\textrm{V}$=1  star. 
This sets the $>100$ second stability requirement for natural guide stars. 
All OPD PSDs are normalized with $\sigma_{10}$ = 10 pm rms. \label{fig:PSDcomp}}
\end{figure}

To understand the impact of wavefront sensing we select a \gls{ZWFS}, which has ideal photon noise limited sensitivity across spatial frequencies \citep{guyon_limits_2005} and is proposed for the baseline \gls{LUVOIR} and the  \gls{HabEx} coronagraph designs \citep{pueyo_luvoir_2017,gaudi_habitable_2018}.
 A \gls{ZWFS} been studied for co-phasing large segmented-aperture space telescopes to  the nanometer level \citep{janin-potiron_fine_2017}, and one is planned for low-order wavefront sensing in the \gls{WFIRST} coronagraph instrument \citep{shi_low_2016}.
 Alternatively, a pyramid wavefront sensor could provide autocalibration of intensity variation, at the expense of increased noise levels \citep{guyon_limits_2005}.
The parameter $\beta_p$ describes the sensitivity of the \gls{WFS} to photon noise for the spatial frequency considered  \cite[Appendix A]{guyon_limits_2005}.
 For a \gls{ZWFS} measuring  rigid-body motion using  photons striking a particular segment  \citep{guyon_limits_2005, ndiaye_calibration_2013},
\begin{equation}
\beta_p = \frac{1}{2}.
\end{equation}
In  Fig.~\ref{fig:PSDcomp} we compare \gls{OPD} \gls{PSD}s with a range of $\alpha$ and $T_0$ values. 
This demonstrates 0.01 Hz is the approximate sensing limit due to stellar photon noise per segment, even for  bright stars (the stellar magnitude limiting photon noise is shown as horizontal solid lines). 
For this discussion, natural guide star  wavefront sensing is limited to the photons solely within V-band \citep{bessell_standard_2005} with the zero-magnitude flux listed in Table \ref{tab:control_params}.
  
\begin{figure}
\centering
\plottwo{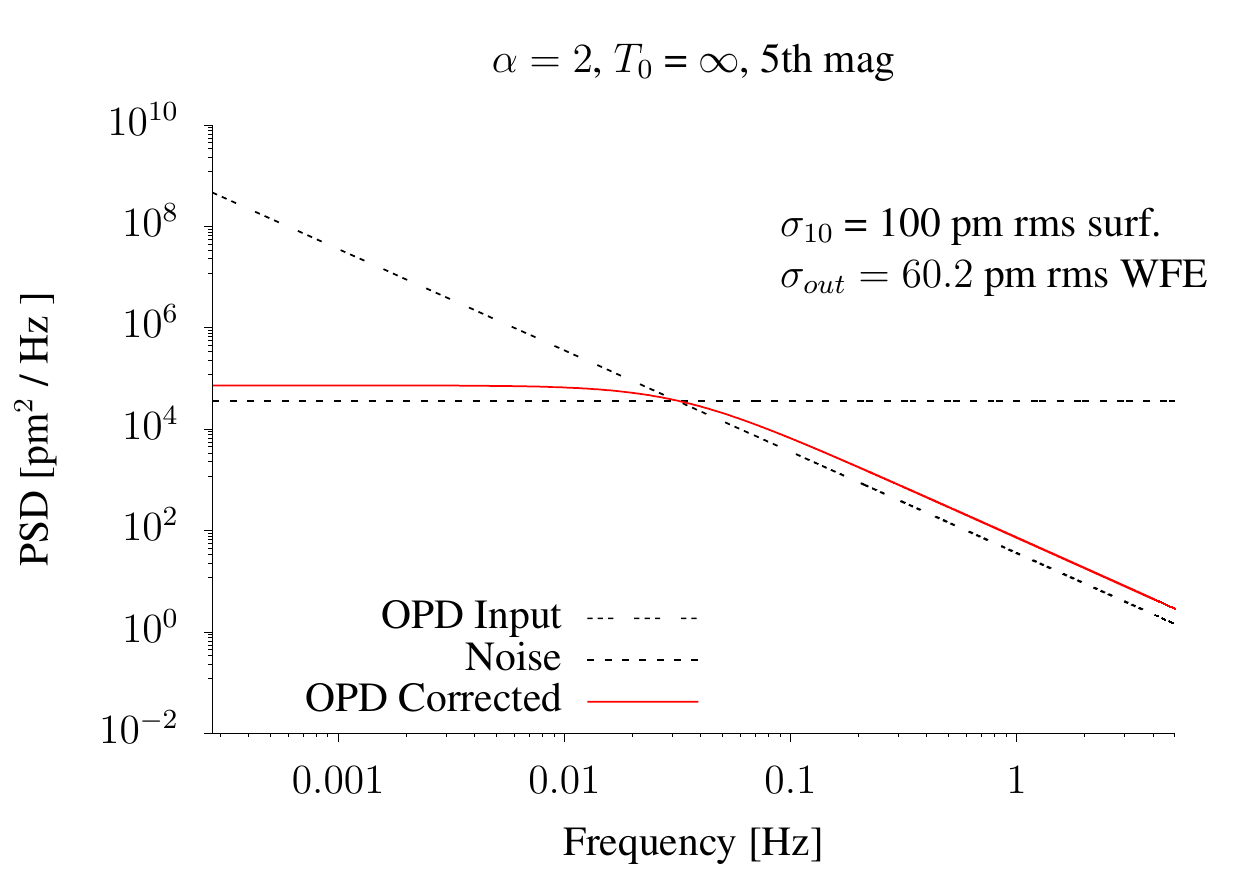}{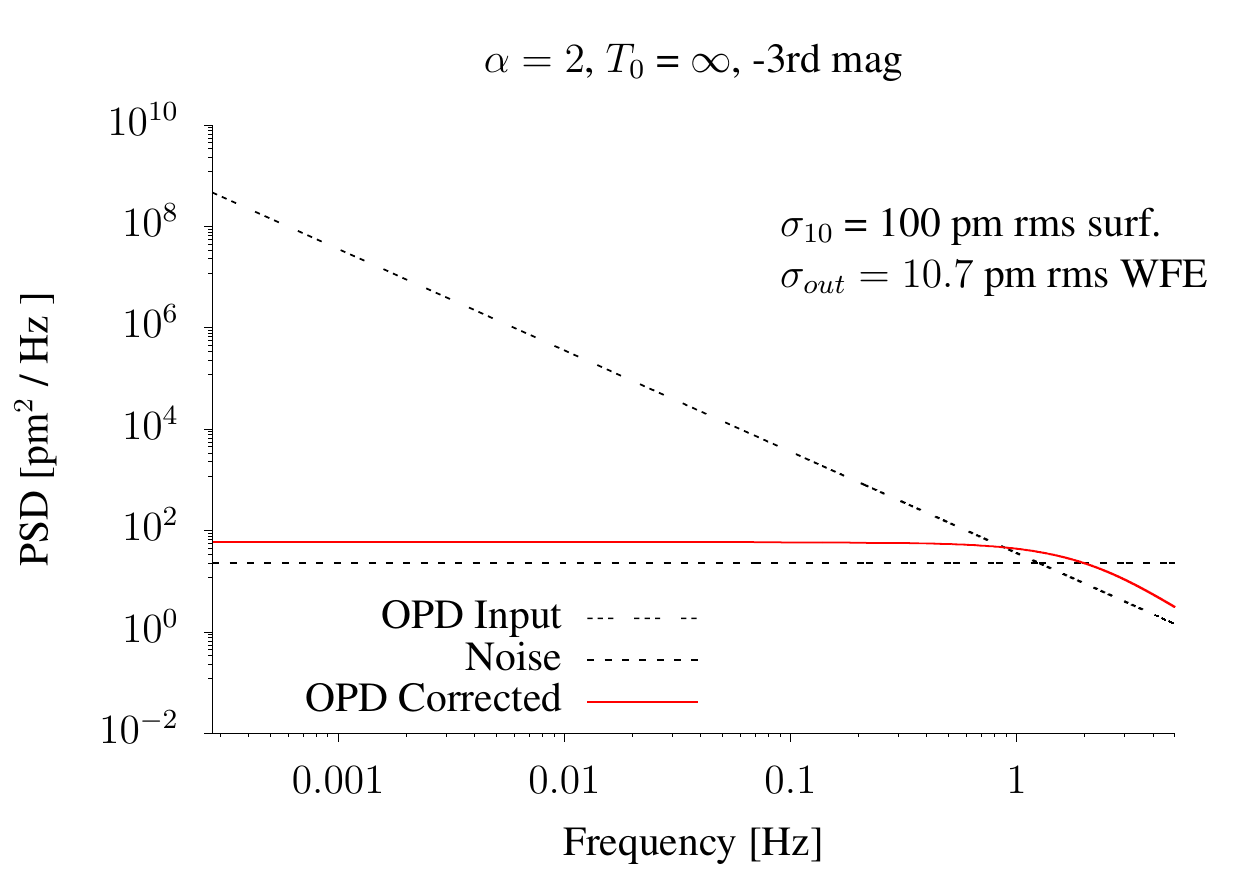}
\caption{Example application of closed loop control to \gls{OPD} using $m_\textrm{V}$=5 (left) and $m_\textrm{V}$=-3 (right) guide stars. 
The input disturbance is shown as a triple dashed line and the corrected output is a solid line.
The photon limited noise floor is shown as a horizontal dashed line. } \label{fig:PSD_control}
\end{figure}

\subsection{Closed Loop Wavefront Control}\label{sec:closed_loop}
In order to assess how close the segment position can be controlled to the stellar sensing limits, we  apply the framework developed in \cite{2018JATIS...4a9001M} for modeling the dynamics of a closed-loop control system.  Given the two PSDs just described, $\mathcal{T}_{OPD}(f)$ (Equation \ref{eqn:vk_psd}) and $\mathcal{T}_{p}(f)$ (Equation \ref{eq:tpf}), the output PSD from a closed-loop control system is given by:
\begin{equation}
\mathcal{T}(f) = \mathcal{T}_{OPD}(f) |ETF(f)|^2 + \mathcal{T}_{P}(f)|NTF(f)|^2,
\end{equation}
where $ETF(f)$ is the system Error Transfer Function and $NTF(f)$ is the system Noise Transfer Function.  These transfer functions describe the action of the control system on the input PSDs, and include the effects of finite integration time, a delay for calculation and communication, and the feedback control law.  
As expected from  Equation \ref{eq:tpf}, the \gls{OPD} contribution of measurement noise is flat versus frequency, while the power law constant (\glssymbol{alpha}) drives the slope of the \gls{OPD} and \glssymbol{T_0} sets the roll off frequency. 
For the example cases where $\alpha=2$ and $T_0=10$ min and  $\alpha=3$ and $T_0=5$, the OPD floor just barely exceeds the noise floor for a 7th magnitude star, so wavefront control on stars dimmer than 7th magnitude would not benefit such a system.

\begin{figure}[htbp]
\begin{center}
\plottwo{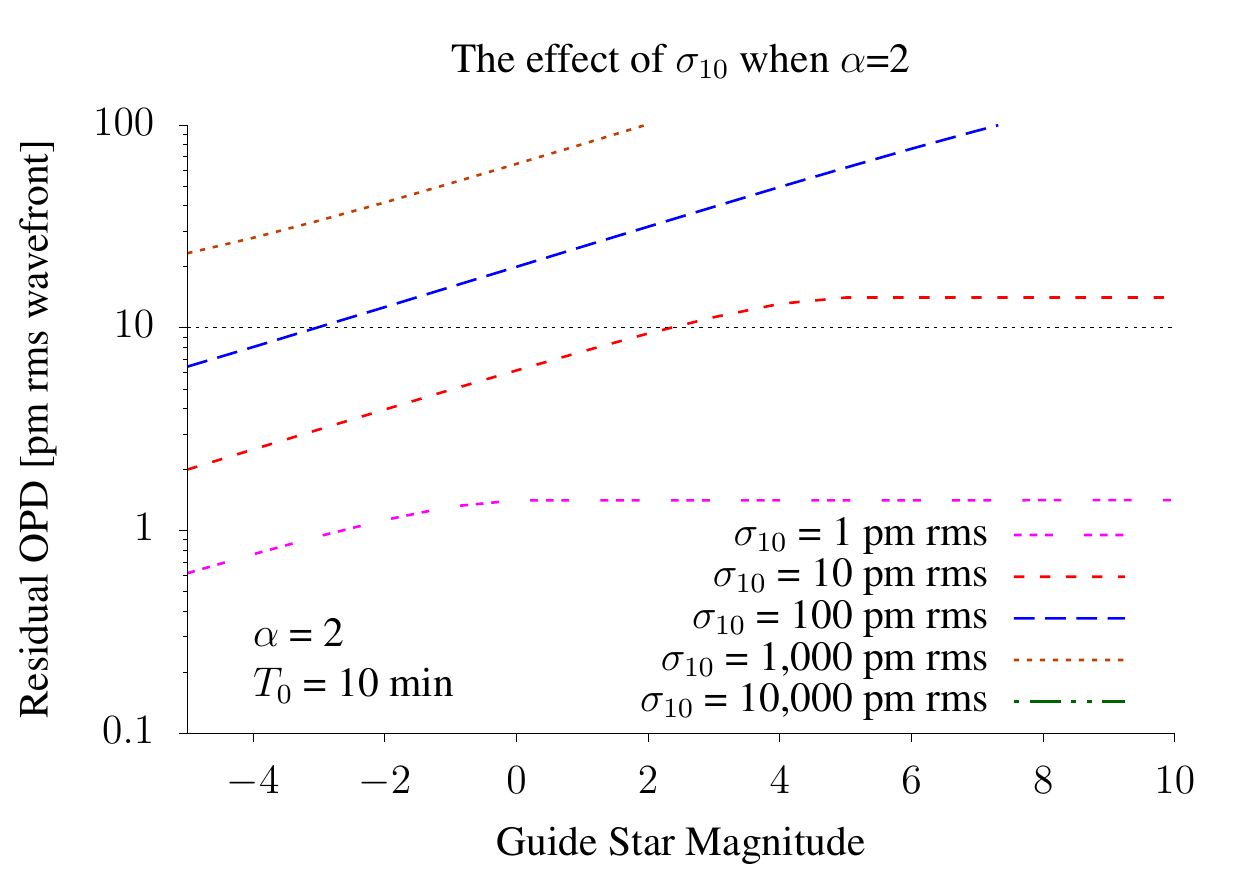}{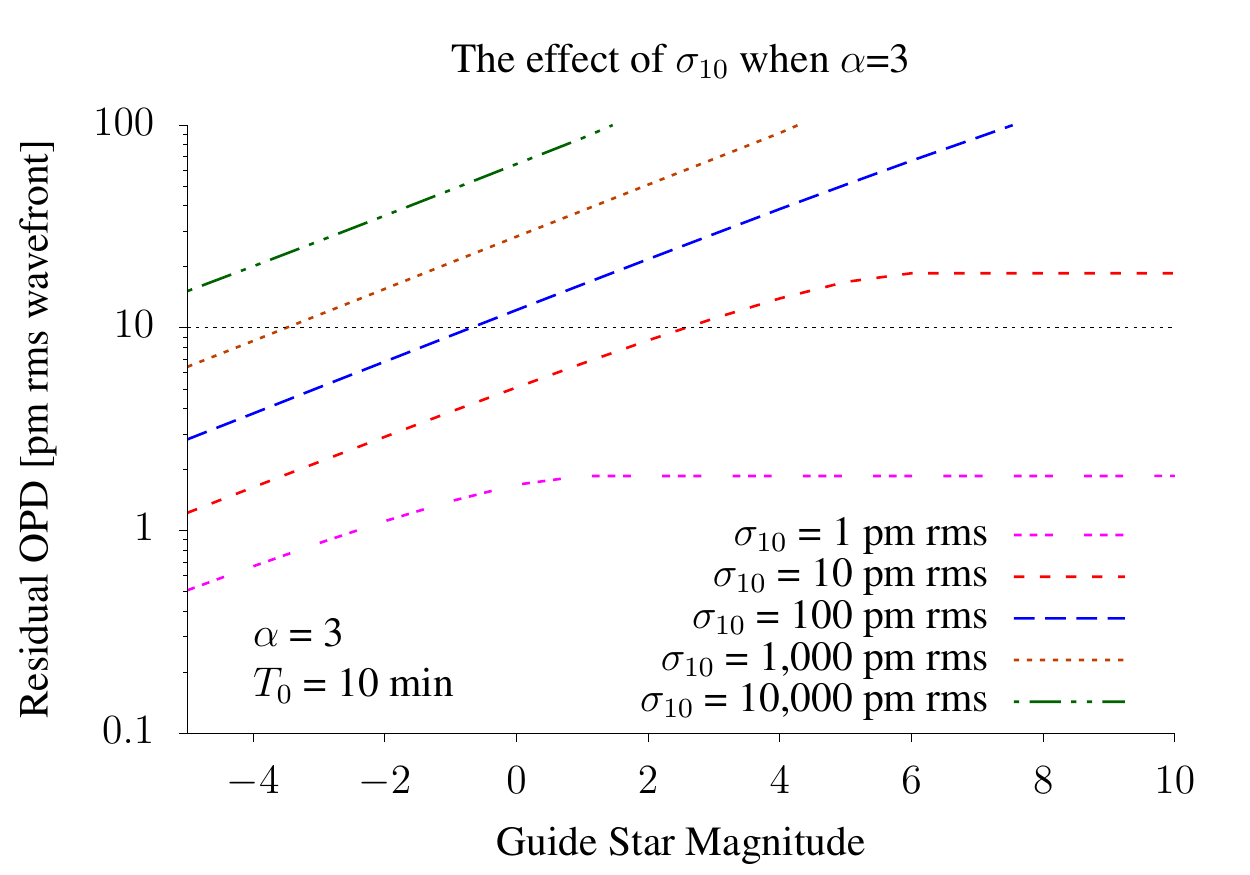}
\caption{{Residual segment disturbance \gls{OPD} as a function of guide star magnitude for $\alpha=2$ (left) and $\alpha=3$ (right) for $T_0=\infty$.
 The \gls{RMS} residual across all temporal frequencies, $\sigma_{out}$, for the  $m_\textrm{V}$=5 star is 60.2 pm which is too high to detect Earth-like exoplanets. 
Conversely, a hypothetical $m_\textrm{V}$=-3 guide star is sufficient to control \gls{OPD} to  10.7 pm and reach contrasts of $\sim10^{-10}$.   
\added{These guide star magnitudes assume a 10\% system throughput. For an ideal detector and minimal loss system with $\sim$40\% throughput, these curves  shift 1.5 magnitudes dimmer.} }
}\label{fig:sigma10compalpha2}
\end{center}
\end{figure}

\section{Artificial Laser Guide Star Spacecraft Concept}
\begin{figure}
    \center
     \includegraphics[width=\textwidth]{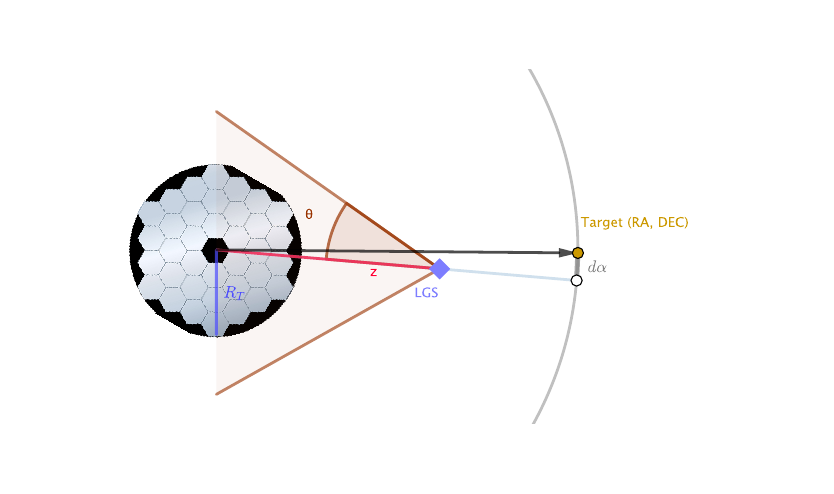}
        \caption{Two dimensional representation of the \gls{LGS} observing scenario. 
                The observing telescope (left) direction of regard is shown as a vector projected onto the celestial sphere (right), while  at range $z$ the \gls{LGS} (tilted square) is offset from the direction of regard by angle $d\alpha$. 
        The width of the Gaussian \gls{LGS}   beam, with a half-width divergence $\theta$, is shaded. 
        }
        \label{fig:coords}
    \end{figure}

Rather than guiding on a science star as assumed previously, one might use an artificial guide star to  achieve  an increased wavefront sensing flux.
In order to maintain 10 pm stability while observing dimmer stars, we explore the potential of a formation-flying spacecraft with a continuous-wave light source, providing more photons than a natural star. 
The geometry of the \gls{LGS} concept is shown in  Fig.~\ref{fig:coords}.
The segmented space telescope with radius $R_T$ is shown at left. 
The \gls{LGS} is shown at a distance $z$, projecting a Gaussian laser beam (to ensure smooth propagation) with a divergence, \glssymbol{theta}, at the telescope. 
While the telescope observes a target star at some astronomical coordinate, the \gls{LGS} appears offset by some angle $d\alpha$.
Table \ref{tab:control_params} includes several key parameters of  the system we will consider for  the design trades throughout this work. 
Section \ref{sec:results} will describe the design constraints on an \gls{LGS} for augmenting an Earth-like exoplanet coronagraph mission using typical telescope properties  drawn from recent publications covering the design of the \gls{LUVOIR} mission concept \citep{pueyo_luvoir_2017,feinberg_ultra-stable_2017}.

\begin{table}
\centering
\caption{System parameters assumed in wavefront sensing and control  calculations. \label{tab:control_params}}
\begin{tabular}{ccl}
Parameter  &  Value & Notes\\
\hline
\hline
Telescope Diameter & 9.2 m & \cite{feinberg_ultra-stable_2017}\\ 
Segment Geometry & Hexagonal &  \cite{eisenhower_atlast_2015}  \\
Segment face-to-face width        & 1.15 m &  LUVOIR A \citep{pueyo_luvoir_2017} \\
Zero-mag photon flux & 9.1$\times10^9$ photons/sec & Vega-based, in Bessel V band\\
System Throughput, \glssymbol{Tp} & 0.1 & including detector QE\\
$\lambda$            & 500 nm &  \\
Loop update rate     & 10 Hz & \\
Loop delay           & 1.5 msec & \\
$\Delta f$           & 1/3600 & The PSDs model 1 hour periods\\
$n_{pix}$            & 16/segment & \\
$\sigma_{ron}$       & 0.3 $e^-$/pixel/frame & typical EMCCD read noise\\ 
\hline
\end{tabular}
\end{table}

\section{Results}\label{sec:results}

\subsection{Photon and Sensor Noise}\label{sec:photonnoise}

The photon noise rate per segment places a limit on the sensing of the segment position.
The closed-loop analysis of Section \ref{sec:closed_loop} relates guide star magnitude, natural or artificial, to residual wavefront error.
 Fig.~\ref{fig:PSD_control} shows the input OPD is well-corrected to near the noise floor at low frequencies.
The input disturbance is shown as a triple-dashed line and is suppressed in the controlled curve (solid line) by more than six orders of magnitude at low frequencies. 
 Wavefront control could be implemented through direct control of segments via a hexapod (e.g. \cite{contos_aligning_2006}), or a deformable mirror (two high-actuator count \gls{MEMS} deformable mirrors are planned for \gls{LUVOIR} \citep{pueyo_luvoir_2017}). 
The most important parameter is the overall level of vibrations, which we have characterized as the 10 minute \gls{RMS}, $\sigma_{10}$. 
The challenge of  controlling these vibrations is shown in  Fig.~\ref{fig:sigma10compalpha2} for the case of $\sigma_{10}$ = 10 min and $\alpha$ = 2 (left panel) and $\alpha=3$ (right panel). For $\sigma_{10}$ = 10 pm, a 10 pm residual \gls{OPD} is achieved for an \added{approximately} 2nd magnitude or brighter guide star star. 
This would limit coronagraphy of a Sun-like stars to within just 3 parsecs for natural guide stars and sets a useful lower limit on the dimmest \gls{LGS} for Earth-like exoplanet imaging. 
\added{The only FGK stars  this nearby and bright are Centauri A and Centauri B; however, as shown in Sec. \ref{sec:discovery_space}, there are dozens of stars of interest of apparent magnitude greater than second and over one hundred which are brighter than third magnitude or have other spectral types.}
To control much larger disturbances, such as $\sigma_{10}$ = 1 nm residual \gls{OPD}, a $m_\textrm{V}$ = -2 or brighter guide star is needed.

\subsection{Pointing Sensitivity}\label{sec:pointing_divergence}
In addition to variations in the pupil plane intensity due to photon noise, changes in the guide star illumination pattern must also be considered in order to set the \gls{LGS} performance requirements.
Unlike the even illumination pattern of a natural guide star, the  \gls{LGS} beam will have a Gaussian intensity distribution.
As discussed Section \ref{sec:photonnoise}, the \gls{ZWFS} is sensitive to both intensity and phase variations. 
The \gls{ZWFS} is effectively an interferometric fringe pattern at a single relative phase shift. Thus, intensity variations  lead to spurious phase measurements. 
This is a concern for the \gls{LGS} because the Gaussian laser beam will produce a variable illumination pattern across the observatory pupil, which moves according to the pointing of the \gls{LGS} relative to the observatory ($d\alpha$ in  Fig.~\ref{fig:coords}).

\begin{figure}
        \includegraphics[width=\textwidth]{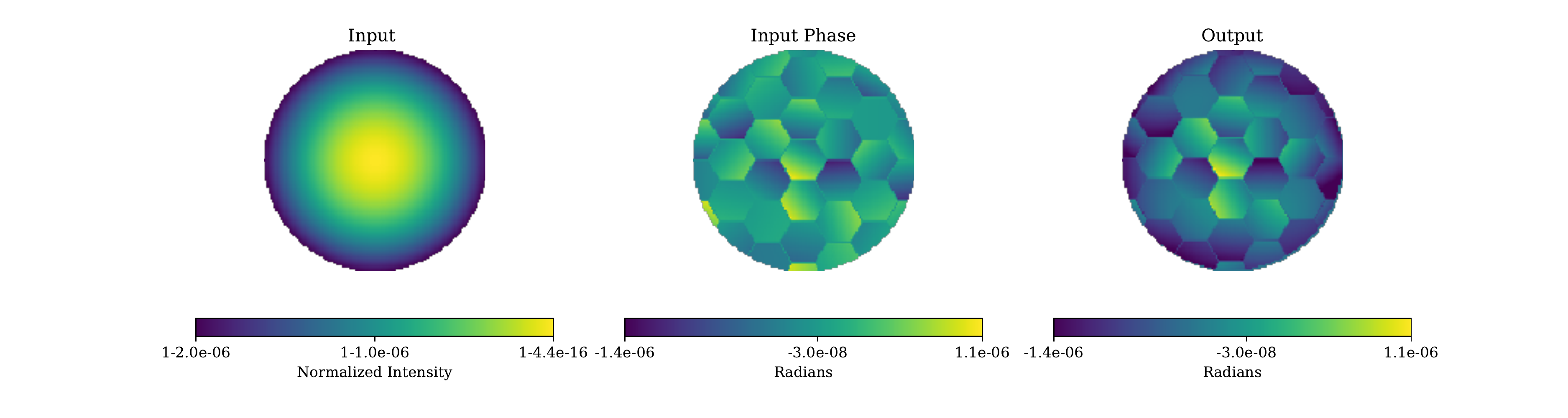}
        \caption{Example of \gls{ZWFS} with  a \gls{LGS}, illustrating that a Gaussian input beam and a flat input phase disturbed by random segment motion (middle) leads to an  varying output intensity (right) which is degenerate with phase errors.  
        Accurate \gls{LGS} pointing allows for quasi-static measurement of the intensity distribution and correction of this error. This simulation used $w=23.40^{\prime\prime}$  at a range of 4.4 $\times10^{4}$ km, which corresponds to a beam-width of 4.7 km at the telescope. }
        \label{fig:ZWFS_flat_phase}
    \end{figure}

In addition to photon noise, if the pupil intensity function is changing on the time scale of segment jitter, due to changes in the pointing of the \gls{LGS}, the \gls{WFS} will sense erroneous tilts across the pupil.
If the Gaussian function is static across the pupil,  then it is straightforward to calibrate a static non-uniform  intensity function. 
For example, Fig.~\ref{fig:ZWFS_flat_phase} shows a simple numerical model of a \gls{ZWFS} generated using the Fresnel propagation environment in the Physical Optics Propagation in PYthon library \cite{perrin_poppy_2016}.
A Gaussian intensity distribution (left panel), and a flat  phase (middle panel) define the input wavefront.
For simplicity and to maintain the optimal choice of mask diameter, the hexagonal aperture was truncated to a circumscribed circle for this  simulation.
After propagation to the image plane and multiplication by a complex phase mask with the optimal diameter \citep{ndiaye_calibration_2013} of 1.06$\lambda$ and $\theta=\pi/2$, propagation to the next pupil plane gives the  measured pupil intensity shown in the right hand panel of Fig.~\ref{fig:ZWFS_flat_phase}.  
For small phase shifts $\phi$,  and an intensity $I$ in a \gls{ZWFS}  pixel,  \cite[Equation 15]{ndiaye_calibration_2013}  
gives the linear relation between phase and intensity as: 
\begin{equation} 
 \phi=I/I_0-0.5. \label{ref:linear_zwfs_phi}
\end{equation}
 $I_0$ is the average intensity across the pupil.
Differentiating shows the phase measurement error as a function of intensity error $dI$ is: 
\begin{equation} 
 d\phi=d I/I_0.
 \end{equation}
 We quantify the pointing jitter error by calculating the fractional intensity difference between an on-axis \gls{LGS} Gaussian beam striking the telescope $I_1'$, and $I_2'$, the same beam offset by \glssymbol{x_0}: 
\begin{equation}
dI/I=\frac{I_1'-I_2'}{I_0}= e^{\frac{-2x^2}{w^2}}-e^{\frac{-2(x+x_0)^2}{w^2}}\label{eq:fractionalIntensyError}
\end{equation}
Where \glssymbol{x} is the radial displacement from gaussian beam, \glssymbol{w} is the beamwidth, and $I_0$ is the intensity. For x=0, then $dI/I'=1-e^{\frac{-2x_0^2}{w^2}}$.
 The first order Taylor expansion is: 
\begin{equation}
d\phi \approx   dI=1-\left(1-\frac{2x_0^2}{w^2}\right)=\frac{2x_0^2}{w^2}.
\end{equation}

Plugging in a minimum required phase error of 10 pm and a pointing error  between measurements (e.g. 15 mas) lets us solve for the minimum beamwidth $w$. 
Due to the need to stabilize intensity across the pupil,  this gives in a ratio of divergence versus transmitter jitter of $>380$, which is much larger the typical values used to maximize received intensity for similar applications such as laser communications   (e.g. ratio of divergence versus transmitter jitter of $\sim 10$, \cite{clements_nanosatellite_2016}).
For \gls{LGS} jitter of 15 mas, a level of performance regularly exceeded 3$\times$ by previous  space observatories (e.g. \cite{nurre_preservicing_1995,koch_kepler_2010,mendillo_flight_2012}), the divergence required to keep the jitter induced errors within 10 pm is $>0.325''$, limited by the \gls{MFD}  (\glssymbol{MFD})  and allows more feasible laser powers.  
\added{Regardless of guide-star approach, for coronagraphic Earthlike-planet imaging, the   observatory pointing jitter must be much lower than that required for the \gls{LGS} transmitter, well below 1 milliarcsecond, in order to maintain $10^{-10}$ contrasts \citep{ruane_performance_2017} .}


\subsubsection{Beam Divergence Limitations}\label{sec:divergence}
The finite size of a single-mode fiber generating the \gls{LGS} beam and diffraction from the exit aperture further limit the minimum  \gls{LGS} beam divergence.
The fiber \gls{MFD} half-angle divergence is given by:
\begin{equation}
\theta_{MFD}=MFD/(2f) 
\end{equation}
where  \glssymbol{MFD} is the mode field diameter of the optical fiber and $f$ is the focal length of the collimating optics. 
In addition to the \gls{MFD},  the size of the exit aperture constrains the beam divergence.
Likewise, for a Gaussian beam, the half-angle beam divergence is given by:
 \begin{equation}
\theta_w=\lambda/(\pi w_0).
\end{equation}
where \glssymbol{w_0} is the beam waist \citep[Equation 22]{kogelnik_laser_1966}.  
To minimize diffraction effects, we assume $w_0$ is one third or less of the \gls{LGS} exit aperture radius. 

\subsection{Wavelength Selection}\label{sec:wavelength}
The \gls{LGS} may contribute background signal to science observations via scattered light, thermal emission, and fluorescence. 
A longer-than-science wavelength out-of-band laser source  minimizes  fluorescence and scattering internal to the telescope (e.g. via dichroic filters)  while a high-efficiency laser minimizes waste heat.
For simplicity, in this initial study, we will consider two common laser wavelengths, 980 nm and 532 nm.
Longer wavelengths allow a decrease in the range between the \gls{LGS} and telescope  (Section \ref{sec:range}), and a 980 nm source is within the sensitivity range of silicon detectors. Efficiency is also critical to designing a  spacecraft with feasible thermal control, and 980 nm lasers have  been previously shown to have provide better than $50\%$  wall-plug efficiency in continuous operation  \citep{crump_optimized_2005}.
Alternatively, the guide laser could be blocked by a narrow line-blocking interferometric filter. \added{Contamination of  high contrast images by \gls{LGS} light presents an additional consideration. Narrow-band interferometric rejection filters with 8 orders of magnitude rejection in \gls{UV} and visible wavelengths have been manufactured \citep{landulfo_novel_2018} and the \gls{LGS} would be further suppressed by keeping the transmitter inside the coronagraph \gls{IWA} (see Sec. \ref{sec:position}).} As described in Section \ref{sec:range},  a shorter wavelength could provide a reference source closer to the center of the visible light science band \cite{pueyo_luvoir_2017}.

\subsection{\gls{LGS} Formation Flying Range}\label{sec:range}

Since the \gls{LGS} is a finite distance from the telescope, we must account for the defocus of the reference wavefront.
For a spherical wave emanating from the \gls{LGS} at distance $z$  (the range to the center of the entrance aperture), the \gls{PV} defocus is given by difference between $z$ and  $R_C$, the range to the edge of the aperture. 
Solving for $R_C$ as a function of telescope aperture and the peak-to-valley wavefront error across the pupil $PV_{WFE}$\added{$=R_C-z$}:
\begin{equation}
R_C=\frac{PV_{WFE}^2+R_T^2}{2PV_{WFE}} \approx \frac{R_T^2}{2PV_{WFE}},
\end{equation}
for a telescope radius $R_T$.
\added{The quasi-linear range where Eq.  \ref{ref:linear_zwfs_phi} holds for a \gls{ZWFS} is approximately $\pm\pi/4$  \citep{ndiaye_calibration_2013}.
Since $\sigma_{10}$<<$\pm\pi/4$, for the configurations considered here, slight variations from non-linearity  are expected to be measurable for calibration}.
Hence, for \gls{PV} wavefront error less than $\pi/2$,
\begin{equation}
R_C=2R_T^2/\lambda. \label{eq:R_C}
\end{equation}
 Thus, the minimum range to the baseline telescope is 43,184 km at 980 nm. 
The addition of a defocusing mechanism in front of the wavefront sensor would relax this requirement, but may add non-common path errors and tighten the lateral stability requirement discussed in Section \ref{sec:position}. 

\subsection{\gls{LGS} position}\label{sec:position}
This section will quantify the station keeping needed, or the accuracy with which the \gls{LGS} must be held on the telescope-target-star vector during a coronagraphic exposure. 
Motion of the \gls{LGS} across the sky relative to the target star will appear as a tilt to a telescope \gls{WFS}. 
For the purpose of this study, we presume the telescope pointing is highly stabilized onboard, such as by a \gls{FGS} \citep{nurre_preservicing_1995}, and that any bulk tilts across the wavefront sensor will be subtracted. 
In order to enable effective tracking of the \gls{LGS}, one might  require it to hold position to within 0.25$\lambda/D$ from the target star ($d\alpha$ in  Fig.~\ref{fig:coords}).
At the baseline range, this corresponds to a cross-track stability of 1 meter, comparable to the requirements of starshade missions (e.g. \cite{soto_starshade_2017}). 
 
\added{Such precise station keeping} would also keep the \gls{LGS} \deleted{behind a coronagraphic mask} \added{inside the coronagraph \gls{IWA}, minimizing contamination from light which leaks past any blocking filters (Sec. \ref{sec:wavelength}). 
Holding the \gls{LGS} on the telescope boresight allows \gls{ADI} and } keeps the wavefront tilt within the range of the \gls{ZWFS} without \deleted{requiring} an additional tip-tilt mirror in the wavefront sensing path, minimizing  sensing of spurious off-axis aberrations. 

Reflected sunlight from the \gls{LGS} could contribute incoherent background to coronagraphic observations.
 For example, neglecting the solar panel reflectivity,  given a 300 mm $\times$ 300 mm spacecraft cross section,  at $5\times10^4$ km range with an an albedo of 0.01, the scattered light is $m_\textrm{V}=16$. 
 A carbon nanotube coated spacecraft could potentially lower the albedo to 0.001, bringing the scattered light as low as $m_\textrm{V}$=19  \citep{cartwright_fifty_2015}.
 This scattered light further motivates keeping the \gls{LGS} well within the inner working angle, as sources inside $\lambda/D$ will be attenuated by many orders of magnitude by most coronagraph designs (e.g. \cite{ndiaye_apodized_2016,trauger_hybrid_2016,zimmerman_shaped_2016})
\added{For example, a charge-6 \gls{VVC} coronagraph suppresses point sources at 0.25$\lambda/D$ by $\sim$ 15 magnitudes \cite{ruane_performance_2017}, making even the brighter scattered light case dimmer than a typical Earthlike planet. 
}


 \subsection{Other considerations}
The \gls{LUVOIR} concept includes simultaneous observations in \gls{UV}, visible, and \gls{IR} with one channel serving as the wavefront sensor \citep{pueyo_luvoir_2017}. 
As with reflected sunlight, discussed in Section \ref{sec:position}, thermal emission of the spacecraft may contribute a significant background if the source is not behind the coronagraph mask.

\added{A variety of means are available to separate incoming light from the \gls{LGS} from the science signal. 
Coronagraph designs, such as \gls{WFIRST}-\gls{CGI} and the \gls{HabEx} \gls{VVC}  use rejected starlight from a reflective focal plane mask at the center of the field to feed a \gls{ZWFS}. Other designs use reflected light from the Lyot stop  \citep{singh_-sky_2015,2015SPIE.9605E..19M}. 
For an \gls{LGS} with precision station keeping (Sec. \ref{sec:position}) both of these approaches, likely in conjunction with a dichroic or notch-blocking filters in the science channel allows separation of \gls{LGS} light from the wavelengths and angles of interest.}
For \gls{LGS} wavelengths shorter than the science wavelength, the magnitude of fluorescence from transmissive optics  \citep{engel_advanced_2003}, and the potential for laser induced contamination of reflecting surfaces \citep{wagner_-situ_2014} will require consideration. 
Fluorescence  effects are dependent on material and wavelength. 
Thus, testing of materials and wavelength selection, along with consideration of a multi-wavelength \gls{LGS} transmitter, is expected to mitigate these effects.

\added{In addition to intensity variations due to pointing jitter, understanding the phase stability of the \gls{LGS} transmitter is critical to assessing feasibility of an \gls{LGS} spacecraft. 
Global changes in the phase due to changes in the lasing wavelength are negligible since the \gls{LGS} is providing a reference wavefront for corrections of relative errors on short time scales. 
Controlling for optical aberrations in a small spacecraft often requires challenging thermal and optical control; however, in this case large aberrations are tolerable.
For a 10 meter-scale telescope, the observatory aperture cross-section is small relative to the range discussed previously, meaning the  error incident on the telescope wavefront is a very small subsample of any internal \gls{LGS} aberrations. 
For example, for the minimum $\theta$ and range in Table \ref{tab:reqs}, the incident beam waist is 0.8 km and a change in the radius of curvature; i.e., due to focus error internal to the \gls{LGS} is small.
To quantify this effect we consider the ratio between the \gls{PV} error across the telescope aperture and across the incident gaussian beam waist $w$, at range $z$, is given by
\begin{equation}
\frac{z-\sqrt{z^2-(D_t/2)^2}}{z-\sqrt{z^2-w^2}}\sim10^{-4},
\end{equation}
effectively minimizing one hundred nanometer scale disturbances across the \gls{LGS} wavefront to picometer scales at the telescope.
Such ``diffraction-limited'' stability is  well within the  range of small satellite optical systems (e.g. \cite{allan_deformable_2018-1}).
}

\section{Discussion}\label{sec:discussion}

There are two key benefits to the LGS approach: the ability to directly image dimmer target star systems, and decreasing the mechanical stability requirements on the telescope.
 Designs at opposite extremes of possible \gls{LGS} transmitted power are shown in Table \ref{tab:reqs}.
 Equation \ref{eq:R_C} sets the range for the mission concepts for two laser wavelengths, 532 nm and 980 nm.
Both rely on an accurately pointed guide star to provide constant intensity calibration where the pupil intensity function is held constant throughout the observation.
For example, a ``well controlled'' 5 W \gls{LGS} case uses a bright guide star which can be sampled quickly at hundreds of Hertz while a "stable" case assumes a relatively stable telescope (e.g. $\sigma_{10}=10$ pm) requiring fewer photons per second for \gls{WFS}ing, is  shown to compare science performance with and without the \gls{LGS}.
The magnitude values in $V$ band are directly comparable to the x-axis of  Fig.~\ref{fig:sigma10compalpha2}, allowing  estimation of the residual \gls{OPD} given a known input \gls{OPD} \gls{PSD}. 
For example, Case VI allows correction of a 1000 pm RMS input OPD to 10 pm RMS for $\alpha=3$.
$z^\prime$ magnitudes are given for the 980 nm \gls{LGS} cases.

\begin{table}
\caption{\gls{LGS} transmitter design parameters possible scenarios where an \gls{LGS} illuminates a segmented telescope. 
The ``stable'' telescope cases, Cases II and IV, are based on recent designs for a well-damped telescope using natural guide stars for wavefront sensing.
The controlled cases, Cases I, III, V, and VI, would use a more powerful laser to enable faster update, relaxing telescope stability requirements.
Throughput here refers to total observatory sensitivity, including the impacts of coating reflectivity and detector quantum efficiency.
Transmitter jitter here is the maximum allowable excursion of the \gls{LGS} beam from the bore-sight of the telescope. 
The rightmost ``band'' column indicates the filter over which the effective magnitude, $m$, is calculated.
}
\begin{centering}
\begin{tabular}{c|c| c|c|c|c|c|c|c|c}\label{tab:reqs}
Case &Telescope & Laser Power & \glssymbol{lambda_wfs} & $z$ & Throughput & $\theta$ & Transmitter jitter &  $m$ & band \\
\hline
 & & $\mathrm{W}$ & $\mathrm{nm}$ & $\mathrm{km}$ &  & $\arcsec$ & $\arcsec$ & $\mathrm{}$ & $\mathrm{}$ \  \\
 \hline
I & Controlled &5.0 & 980.0 & 43184.0 & 0.1 & 3.861 & 0.015 & -7.1 & $\textrm{z'}$ \\
II& Stable Telescope  & 0.005 & 980.0 & 43184.0 & 0.1 & 3.861 & 0.015 & 0.4 & $\textrm{z'}$ \\
III& Controlled & 5.0 & 532.0 & 79549.0 & 0.1 & 3.861 & 0.015 & -4.9 & $\textrm{V}$ \\
IV& Stable Telescope & 0.005 & 532.0 & 79549.0 & 0.1 & 3.861 & 0.015 & 2.6 & $\textrm{V}$ \\
V& Controlled & 5.0 & 980.0 & 43184.0 & 0.1 & 12.49 & 0.1 & -4.5 & $\textrm{z'}$ \\
VI& Controlled & 5.0 & 532.0 & 79549.0 & 0.1 & 12.49 & 0.1 & -2.3 & $\textrm{V}$ \\
\end{tabular}
\end{centering}
\end{table}

\subsection{Controlled Case: Relaxed Telescope Stability}\label{sect:relaxed}
In addition to increasing the available discovery space, an \gls{LGS} has the potential to drastically relax telescope stability requirements\added{, potentially decreasing spacecraft mass and cost}.
Since the system control authority is no longer limited by photon noise, primary mirror segments can be actively held in position.
For example,  Fig.~\ref{fig:sigma10compalpha2} shows that, for $\alpha=3$, a $m_\textrm{V}$ = -4 or dimmer  guide star (cases I,  III,  and V in Table \ref{tab:reqs}) could provide 10 pm segment control for a $\sigma_{10}$ = 1 nm input disturbance when $T_0=$ 10 min.  
This is a more than two orders of magnitude of relaxation in telescope stability compared to performing wavefront sensing  using photons from a $m_\textrm{V}$ = 5 science target.  
Alternatively, a shorter $T_0$ could be controlled with a brighter \gls{LGS} or a smaller $\sigma_{10}$.

\subsection{Stable Case:  Increased Discovery Space}\label{sec:discovery_space}

It is illustrative to estimate how reducing wavefront sensing noise impacts science yield for a telescope built with sufficient segment stability to reach $10^{-10}$ contrasts around $m_\textrm{V}$ = 5 stars.

A highly corrected telescope coupled with an \gls{LGS} would open up a large population of nearby candidate host stars to greater than $10^{-10}$ contrast imaging with future space telescopes with a low-power $\lesssim$50 mW \gls{LGS} transmitter.
This scenario for 532 nm and 980 nm laser wavelengths is shown in cases II and IV in Table \ref{tab:reqs}.
These stars may not be ideal for uninformed searches due to the longer exposure times required, but if more rocky exoplanets are discovered by upcoming sub-1 m/s radial velocity surveys of stars as dim as V=12 (e.g. \cite{halverson_comprehensive_2016}), the capacity for high-contrast imaging of stars with low apparent magnitude will be valuable.

For a well stabilized telescope, the transmitter jitter requirement for the \gls{LGS} could be relaxed from the 15 mas discussed in Section \ref{sec:pointing_divergence}. 
Given the divergence and pointing constraints discussed in Section \ref{sec:pointing_divergence}, Case V and VI show that example systems with a relaxed transmitter jitter of 0.1$^{\prime\prime}$ and correspondingly increased $\theta$ still result in a guide star that is several magnitudes brighter than science targets, providing several magnitudes of margin on 10 pm segment rigid-body sensing.

    \begin{figure}
    \center
      \includegraphics[width=0.5\textwidth]{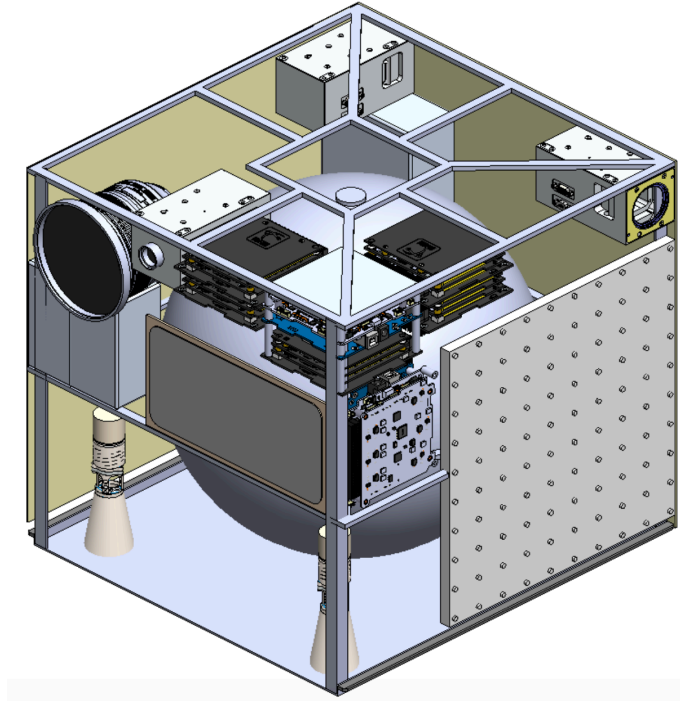}
        \caption{Rendering of a 30 cm $\times$ 30 cm $\times$ 30 cm (27U)  small satellite guide star built around a spherical fuel tank.
        A notional laser transmitter is shown in the  upper left. 
        Low reflectivity paneling and photovoltaic panels are not shown. }
        \label{fig:27U_LAYOUT_LABELS}
    \end{figure}

Previous research has suggested that 10 minute stability at the 10 picometer level is necessary for the detection of Earth-like planets around 5th to 6th magnitude stars  \citep{stahl_engineering_2013,stahl_preliminary_2015}.
The analysis presented here shows that the combination of a limiting magnitude of $m_\textrm{V}$ = 5 to $m_\textrm{V}$ = 6 and $\sigma_{10}$ = 10 minutes may be overly optimistic.
The ideal coronagraph model output shown in  Fig.~\ref{fig:contrast_vs_disturbance} supports the previous finding that 10 pm control of segment motion is necessary to reach contrasts of $10^{-10}$ at the \gls{IWA}. However, after accounting for system transmission, detector noise, and photon noise using a closed-loop control law,  Fig.~\ref{fig:sigma10compalpha2} shows controlling segment rigid-body motion to this level requires guide star magnitudes $m_\textrm{V}$ < $3$.

\begin{figure}[t]
\begin{center}
     \includegraphics[width=\textwidth]{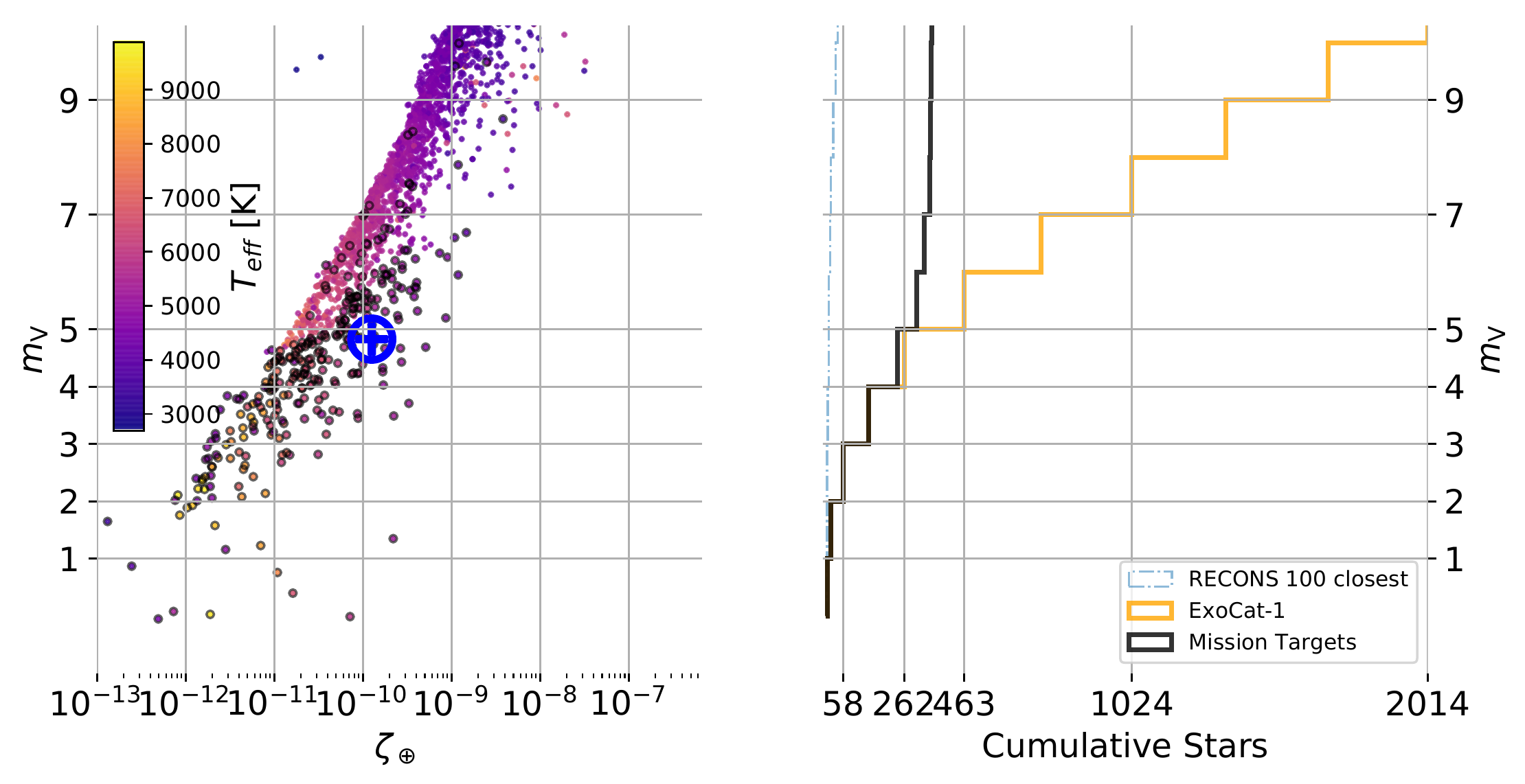}
\caption{Left: Magnitude vs flux ratio of a 1$R_\oplus$ planet at the Earth-equivalent insolation distance for stars in the ExoCat database below 10,000 Kelvin \citep{turnbull_exocat-1_2015}. 
The sun at 10 parsecs is shown as the encircled cross.
Black outlined stars have been previously flagged by NASA missions as priorities for the \gls{WFIRST} \citep{kasdin_wfirst_2018}, \gls{LBTI} \citep{ertel_hosts_2018},  Exo-S \citep{shaklan_exo-s_2015}, and Exo-C \citep{stapelfeldt_exo-c_2015} projects in the Exoplanet Archive.
Right: The cumulative number of target stars as function of magnitude.   There are 40 stars brighter than 10th magnitude in the RECONS databases of the closest stars and 2,014 in ExoCat. 
Natural guide star wavefront sensing and 10 minute telescope stability place a limit at 3th magnitude or brighter which leaves 142 ExoCat stars accessible. 
The solid black line shows the same mission stars as are circled in the left panel.}
\label{fig:vmag_vs_c}
\end{center}
\end{figure}

Unfortunately, many  promising, nearby, exoplanet host stars are dimmer than either 3rd or 5th magnitude, particularly M-dwarf stars, which may host the majority of terrestrial planets \citep{dressing_occurrence_2015,shields_habitability_2016}. 
While the habitable zone is still largely unconstrained (see discussion in \cite{seager_exoplanet_2013}); we explore the flux ratio of a canonical Earth-radius planet ($R_\oplus$) at the Earth equivalent insolation distance, $r_{eei}$.
This flux ratio is given by  
\begin{equation}
\zeta_\oplus=\frac{A \Theta(\alpha)R^2_\oplus}{ r^2_{eei}}.
\end{equation}
 Fig.~\ref{fig:vmag_vs_c} plots $m_\textrm{V}$ versus $\zeta_\oplus$ using data from ExoCat-1 \citep{turnbull_exocat-1_2015}, for planets at the Earth-equivalent insolation distance with geometric albedo $A= 0.2$ and a typical $\Theta=1/\pi$ phase function value \citep{robinson_characterizing_2015}.
 These targets can be broken into four quadrants around the  canonical Earth-Sun value at 10 pc, indicated with an encircled cross just above $10^{-10}$ at slightly below $m_\textrm{V}=5$. 
Below 10$^{-10}$ are hotter stars with habitable zones at farther separations. 
Dimmer than 5th magnitude, the majority of target stars are cooler than the Sun with slightly larger $\zeta_\oplus$ values, but there is a significant population of Sunlike and hotter  stars within 30 parsecs with lower $\zeta_\oplus$ values which will be particularly hard to access with natural guide star sensing.
The right panel of Fig.~\ref{fig:vmag_vs_c} shows the cumulative distribution versus $m_\textrm{V}$ for the \gls{RECONS}\footnote{\url{http://www.recons.org/TOP100.posted.htm}, last updated 2012 Jan 1.}  nearest 100 stars, and  the EXOCAT-1 catalog of promising exoplanet host stars \citep{turnbull_exocat-1_2015}. 
 There are 463 exoplanet candidate host stars with $m_\textrm{V} \leq$ 5 within 30 pc in the ExoCat-1 catalog and more than four times as many stars if the limiting magnitude is instead extended to 10th magnitude.
 As can be seen from the shading of points in  Fig.~\ref{fig:vmag_vs_c}, many of these dim stars are cooler than the Sun, with less stringent contrast requirements.
There are hundreds more stars between 5th and 8th magnitude with comparable flux ratios to the Earth at 1 AU from the Sun.
For the more conservative $m_\textrm{V}<3$ limit found here, there are only 142 target stars visible. 
An \gls{LGS} allows recovery of the contrast needed to search the future mission target stars for Earth-like planets with a coronagraph. 
Alternatively, a telescope stability, $\sigma_{10}$, $<10$ pm or a stability outer time, $T_O$, $>$10 minutes is needed.


\section{Summary}\label{sec:summary}

This paper is intended to serve as a starting point in the explorations of design space for a \gls{LGS} to control segmented telescope motion.
We have summarized the key design parameters, including inter-spacecraft range, source wavelength, station keeping, and laser transmitter jitter for a spacecraft formation flying along the direction  of regard of a segmented-aperture coronagraphic space telescope to provide a bright reference wavefront.
 By applying a closed-loop transfer function to  wavefront control of rigid-body motion for the nominal \gls{LUVOIR} segment geometry, we derive a wavefront sensing limiting magnitude for detection of Earth-like planets of $m_\textrm{V}<3$ with an ideal coronagraph, given 10 minute telescope stability.
The \gls{LGS} concept as described enables relaxation of telescope segment  stability by up to two orders of magnitude and offers the potential for nearly an order of magnitude increase in the number stars observable to $10^{-10}$ raw contrast without wavefront sensing limitations.

This work will be followed by more detailed studies to optimize \gls{LGS} feasibility and performance.
Areas where further design studies are required include the number and lifetime of \gls{LGS} spacecraft, range compensating wavefront sensor fore-optics,  and alternative wavefront sensor architectures. 
In particular, laser wavelengths could be significantly shorter or longer; either increasing the sensitivity per photon or decreasing the wavefront curvature at a given range.
Decreasing curvature would allow the \gls{LGS} to fly closer and decrease maneuvering costs.
It may be possible to trade  these notional requirements for increased system complexity. 
For example, a focus and pointing correction stage could allow shorter \gls{LGS}-telescope separations and relaxed station keeping requirements.
Studies and laboratory simulations of the non-common-path ray propagation and higher-order diffraction effects of such solutions are presently underway (Xin et al.~ in preparation, and \citep{lumbres_modeling_2018-1}.

Large disturbances with steep power law distributions, $\alpha>3$,  are readily correctable by a \gls{LGS} of feasible brightness, which potentially enables relaxed segment positional stability requirements, decreasing the engineering changes relative to the structural design of the \gls{jwst}.
Predictive control \citep{males_ground-based_2018} could also improve performance  at resonance  frequencies. For example, \gls{jwst}  has a 20 nm \gls{RMS} segment ``rocking'' mode  at $\sim$40 Hz \citep{stahl_preliminary_2015}.  
The results above show care must be taken to specify the full \gls{PSD} envelope of \gls{OPD} disturbances, otherwise the actual limiting magnitude may be much brighter than expected.

The details of the \gls{LGS} transmitter needed to provide precision pointing, particularly whether it is stabilized by a fine pointing system or body pointing, are  the subject of future work along with development of control laws and quantification of the noise requirements for the attitude sensors and actuators.

Efforts are underway (captured in Clark et al., in preparation) to also develop mission architectures and spacecraft designs that optimize in terms of terrestrial planet yield while integrating state-of-the art power, thermal, and propulsion technologies.
\added{These efforts include detailed operational restrictions due to reflections and  orbital requirements at the Sun-Earth Lagrange Point 2 (L2)}.
 Coordination between multiple \gls{LGS} in order to minimize delay between observations could significantly increase observing efficiency; a similar approach has been proposed for starshades, which have more complex systems with stringent requirements on fabrication, deployment, attitude, and navigation \citep{stark_maximized_2016}.

\acknowledgments
This work was made possible by a NASA Early Stage Innovation Award, \#NNX17AD07G.
The team is grateful to Lee Feinberg and Ian Crossfield for many useful conversations.
\added{The authors would also like to thank the reviewer for many helpful comments}.
This research has made use of the NASA Exoplanet Archive\footnote{accessed 16 July 2018}, which is operated by the California Institute of Technology, under contract with the National Aeronautics and Space Administration under the Exoplanet Exploration Program.
This research has made use of the SIMBAD database, operated at CDS, Strasbourg, France.
\added{This research has made use of  data from \gls{RECONS} \citep{henry_solar_2018}. 
Figures related to \gls{ZWFS} and yield calculations, as well as the numerical calculations in this work are available as a Jupyter notebook \citep{douglas_douglase/aj_lgs_2018_plotsandfigs_2018}.}

.
\facilities{Exoplanet Archive} 
\software{This research made use of community-developed core Python packages, including: Astroquery \citep{adam_ginsburg_astropy/astroquery_2018}, Astropy \citep{the_astropy_collaboration_astropy_2013}, Matplotlib \citep{hunter_matplotlib_2007}, SciPy \citep{jones_scipy_2001}, and
the IPython Interactive Computing architecture \citep{perez_ipython_2007}.}
 
\clearpage 
\bibliographystyle{yahapj}
\bibliography{atlast,report}

\end{document}